\documentclass[superscriptaddress,pra,twocolumn,showpacs,preprintnumbers,amsmath,amssymb,floatfix]{revtex4}

\usepackage{graphicx}
\usepackage{dcolumn}
\usepackage{bm}
\usepackage[mathscr]{eucal}
\usepackage[nooneline]{subfigure}

\newcommand{\etal}{\textit{et al}}

\newcommand{\bra}[1]{\langle#1\vert}
\newcommand{\ket}[1]{\vert#1\rangle}
\newcommand{\op}[2]{\ket{#1}\bra{#2}}
\newcommand{\inner}[2]{\bra{#1} #2\rangle}
\newcommand{\braket}[1]{\langle #1 \rangle}

\newcommand{\eqn}[1]{Eq.~(\ref{#1})}
\newcommand{\eqns}[1]{Eqs.~(\ref{#1})}
\newcommand{\ignore}[1]{}

\newcommand{\basisket}[3]{\ket{#1}\ket{#2}\ket{#3}}
\newcommand{\basisbra}[3]{\bra{#1}\bra{#2}\bra{#3}}
\newcommand{\commute}[2]{\left[#1,#2\right]}

\newcommand{\T}{\mathcal{T}}
\DeclareMathOperator{\Tr}{Tr}
\newcommand{\tr}[1]{\Tr\left\{#1\right\}}
\newcommand{\fig}[1]{Fig.~\ref{#1}}
\newcommand{\figs}[1]{Figs.~\ref{#1}}

\newcommand{\jump}{{\mathcal J}}

\newcommand{\x}{\mathsf{x}}
\newcommand{\y}{\mathsf{y}}
\newcommand{\poli}{\mathsf{i}}
\newcommand{\polj}{\mathsf{j}}
\newcommand{\bx}{{XX}}
\newcommand{\xx}{X\mathsf{x}}
\newcommand{\xy}{X\mathsf{y}}
\newcommand{\g}{\mathsf{G}}
\newcommand{\hc}{\mathsf{H.c.}}
\newcommand{\n}{\hat n}
\newcommand{\nxa}{\n_{\x,\omega_A}}
\newcommand{\nya}{\n_{\y,\omega_A}}
\newcommand{\nxb}{\n_{\x,\omega_B}}
\newcommand{\nyb}{\n_{\y,\omega_B}}
\newcommand{\Ax}{a_{\x,\omega_A}}
\newcommand{\Ai}{a_{\poli,\omega_A}}
\newcommand{\Ay}{a_{\y,\omega_A}}
\newcommand{\Bx}{a_{\x,\omega_B}}
\newcommand{\Bi}{a_{\poli,\omega_B}}
\newcommand{\By}{a_{\y,\omega_B}}
\newcommand{\vis}{\mathcal{V}}
\newcommand{\Bell}{\mathcal{B}}
\newcommand{\meanBell}{\bar \Bell}
\newcommand{\basis}{B}
\newcommand{\prob}{\mathcal{P}}
\newcommand{\Q}{\mathcal{Q}}
\newcommand{\state}{\tilde \psi}
\newcommand{\out}{\mathrm{out}}
\newcommand{\inop}{\mathrm{in}}
\newcommand{\meanvis}{\bar \vis}
\newcommand{\excitationnumber}{\hat N}
\newcommand{\subspace}[1]{\mathcal{S}_{#1}}
\newcommand{\heff}{\mathcal{H}}

\begin{document}


\title{An entangled two photon source using biexciton emission of an asymmetric quantum dot in a cavity}

\author{T.M. Stace}
\affiliation{Cavendish Laboratory, University of Cambridge,  Madingley Road, Cambridge CB3 0HE, UK}
\email{tms29@cam.ac.uk}
\author{G.J. Milburn}
\affiliation{DAMTP, University of Cambridge,  Wilberforce Road, Cambridge CB3 OWA, UK}
\affiliation{Centre for Quantum Computer Technology,
University of Queensland, St Lucia, QLD 4072, Australia}
\author{C.H.W. Barnes}
\affiliation{Cavendish Laboratory, University of Cambridge,  Madingley Road, Cambridge CB3 0HE, UK}

\date{\today}



\begin{abstract}
A semiconductor based scheme has been proposed for generating entangled photon pairs from the radiative decay of an electrically-pumped biexciton in a quantum dot.  Symmetric dots produce polarisation entanglement, but experimentally-realised asymmetric dots produce photons entangled in both polarisation and frequency.
In this work, we investigate the possibility of erasing the `which-path' information contained in the frequencies of the photons produced by asymmetric quantum dots to recover polarisation-entangled photons.  We consider a biexciton with non-degenerate intermediate excitonic states in a leaky optical cavity with pairs of degenerate cavity modes close to the non-degenerate exciton transition frequencies.  An open quantum system approach is used to compute the polarisation entanglement of the two-photon state after it escapes from the cavity, measured by the visibility of two-photon interference fringes.  We explicitly relate the two-photon visibility to the degree of Bell-inequality violation, deriving  a threshold at which Bell-inequality violations will be observed.  Our results show that an ideal cavity will produce maximally polarisation-entangled photon pairs, and even a non-ideal cavity will produce partially entangled photon pairs capable of violating a Bell-inequality.
\end{abstract}
\maketitle
\section{Introduction} \label{Intro}

Recent proposals for quantum communication \cite{nie00,gis02} and quantum 
information protocols \cite{kni01} provide a significant incentive to develop practical single photon sources and entangled two photon sources.
The first requirement for such sources is that the emission times of the 
photons be periodic with a precisely defined clocked frequency. Exciton recombination in electrically or optically
excited quantum dots is a candidate system for such sources. In this paper we will discuss an entangled two photon source based on 
recent experiments in self assembled interface quantum dots \cite{yua02,shi02}.
A proposal for producing entangled photon-pairs on demand based on biexciton emission 
from a quantum dot was recently presented by Benson et al. \cite{ben00}.  

A pair of excitons confined in a quantum dot form a bound state known as a biexciton. 
The decay of the biexciton proceeds by consecutive single
electron hole recombination processes. This is estabished experimentally by
the temporal correlation of the biexciton emission and the exciton emission; 
time resolved photoluminescence measurements show the 
exciton photon to be emitted {\em after} the biexciton photon \cite{shi02}. A similar time resolved study of the 
polarisation of the emitted photons shows that there are two decay paths, and it has been shown that they are coherent with one another \cite{che02}. 
While the biexciton photon and the exciton photon emitted in each decay path have the same linear polarisation, 
the polaristion in different decay paths are orthogonal. If these decay paths were indistinguishable then 
this would be a good candidate for an entangled two photon source. Unfortunately small asymmetries in the 
physical geomery of the dots makes the two paths distinguishable, since the asymmetry of the dot breaks the degeneracy of an intermediate 
exciton level enabling the two paths to be distinguished  by frequency.  The effect of asymmetry on the spectrum of  excitons in dots was observed experimentally in dots formed by monolayer fluctuations in a GaAs 2D quantum well \cite{gam96} and has been addressed theoretically \cite{gup98}.  It has also been observed experimentally in CdSe/ZnSe dots \cite{kul99} and in self assembled GaAs/InGaAs dots \cite{ste02}. 
In \fig{fig:EnergyLevels} we indicate the possible decay paths 
from a single biexciton level  through two non degenerate exciton levels to the 
ground state of the dot. The first decay path corresponds to the
emission of a biexciton photon with linear polarisation in the $\x$-direction 
at frequency $\omega_1$, followed by the emission of the exciton photon,
with the same polarisation, at frequency $\omega_2$. In the second decay 
path the biexciton emits a $\y$-polarised photon at frequency $\omega_3$
followed by the exciton emission, also with $\y$-polarisation, at frequency $\omega_4$. 

The state of the emitted photon pairs may then be written as
\begin{equation}\label{eqn:frequency_entangled}
\ket{\psi_{1}}=(\ket{\x,\omega_1;\x,\omega_2}+\ket{\y,\omega_3;\y,\omega_4})/\sqrt{2},
\end{equation} 
where the notation indicates the mode (polarisation and frequency) occupied by each photon of the pair, $\ket{\textrm{photon 1};\textrm{photon 2}}$, with the order reflecting the order of emission. It has been established experimentally that the weights of the kets are equal \cite{ste02}.  In contrast, we wish to produce a state of the form
\begin{equation} \label{eqn:polarisation_entangled}
\ket{\psi_{2}}=\ket{\x,\omega_A;\x,\omega_B}+\ket{\y,\omega_A;\y,\omega_B}/\sqrt{2},
\end{equation}
and we call such a state polarisation entangled, since the entanglement is only in the polarisation degree of freedom.  This is in contrast to the state given by \eqn{eqn:frequency_entangled} which is entangled in both polarisation and frequency.  The important difference between states $\ket{\psi_{1}}$ and $\ket{\psi_{2}}$ is that the second ket in $\ket{\psi_{2}}$ may be rotated into the first ket using linear optical elements such as half-wave plates (HWP) and polarising beam splitters (PBS), and vice-versa, whereas this is not possible for the two kets written in state $\ket{\psi_{1}}$.  Thus, for instance, Bell-inequality measurements and two-photon interference experiments may be performed with realtive ease using $\ket{\psi_{1}}$ but not $\ket{\psi_{2}}$, and this translates to a technological setting in, for instance, quantum key distribution.


The problem of producing frequency-and-polarisation entangled states akin to $\ket{\psi_1}$ has been considered for photon pairs produced by spontaneous parametric down-conversion
in a nonlinear crystal \cite{bra99}.  In this case photons are also entangled both in polarisation and frequency, though the frequency entanglement is more complicated.  The frequencies for the two emitted photons are constrained by energy conservation, so that their sum must equal the frequency of the absorbed pump photon.  Since this single constraint does notdetermine the frequencies of the two emitted photons uniquely, each photon of the pair may be emitted over a wide range of frequencies determined by the spectrum of the pump pulse and the phase matching requirement (which is an expression of momentum conservation).  Thus the photon pair is entangled in its frequency degree of freedom. 

A resolution to this problem, presented and experimentally implemented in \cite{bra99}, is to
pass the signal and idler beams back through the crystal, but with the polarisations rotated through $\pi/2$, with the result that the two ways in which the photons can be emitted with correlated polarisation are not distinguished by frequency.
This scheme in \cite{bra99} does not directly translate to the case of biexcitonic emission, but we nevertheless wish to remove the spectral dependence from the entanglement in state $\ket{\psi_1}$, so the aim of this paper is to present and analyse a proposal to accomplish this for the biexciton entangled photon source.

In this paper we demonstrate that the frequency may be disentangled from the polarisation by placing the dot in an external cavity with suitably chosen cavity-exciton coupling strengths and cavity mode frequencies.  We will show that the external cavity can erase the ``which-path" information contained in the frequency components of state $\ket{\psi_1}$.  The external cavity is used to control both the spectral and spatial mode structure of the emitted photons to enable  the entanglement  to be demonstrated in an interferometer.  A similar
idea using waveguides for SPDC has been proposed by Banaszek \etal\ \cite{ban01}.  We note that the original proposal for the two-photon source \cite{ben00} includes the external cavity, but its presence is only to increase the outcoupling efficiency, and only brief mention is made of its effect upon the spectral emission properties of the emitted photons.

The next part of this paper begins by defining a Hamiltonian for a four level system interacting with optical cavity modes. A master equation is developed in the third section to deal with photons leaking from the cavity and into some measurement apparatus, as well as to account for decoherence events such as photon loss.  In the fourth section, we discuss some operational definitions to quantify the entanglement of the photons produced, such as two-photon visibility and Bell-inequality violations, with the aid of which we judge the efficacy of the cavity in restoring the polarisation entanglement.  We then provide some results in the fifth and sixth sections showing that an ideal cavity does establish maximally entangled photon pairs, and numerical results showing how sensitive the resulting state is to imperfections in the system parameters.  We then provide some heuristic analytic results in the discussion which explain the numerical results, as well as comment on implications for experiments, and finally conclude the paper.

\section{System Hamiltonian}

\label{sec:Hamiltonian}
\begin{figure*}
\subfigure[]{\includegraphics{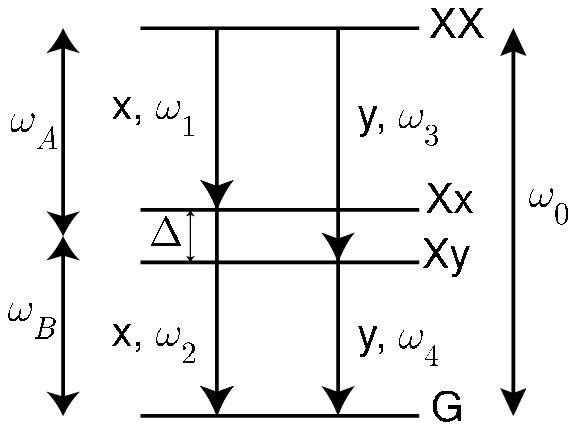}\label{fig:EnergyLevels}}\hfill
\subfigure[]{\includegraphics{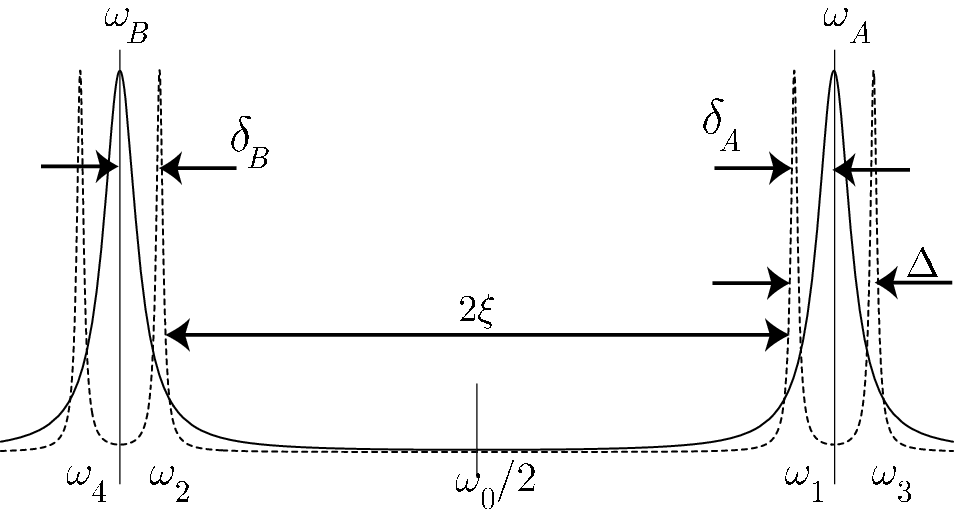}\label{fig:Spectrum}}
\caption{
(a) Energy level diagram and available transitions for the quantum dot and cavity system.  (b)  Spectrum of exciton transitions (dotted line) and cavity modes (solid line) indicating the relevant frequencies for the interaction Hamiltonian.}
\end{figure*}

Figure \ref{fig:EnergyLevels} shows the energy levels and available dipole transitions for the biexciton-cavity system.  The biexciton state is given by $\ket{\bx}$, $\ket{\xx}$ and $\ket{\xy}$ are the intermediate excitonic states in the $\x$ and $\y$ polarisation decay paths respectively and $\ket{\g}$ is the dot ground state.  The cavity is assumed to support pairs of degenerate $\x$- and $\y$-polarised modes at frequencies $\omega_A$ and $\omega_B$.  In our model, we do not include coupling between for instance the cavity mode $\ket{\omega_A,\x}$ and the transition $\ket{\g}\leftrightarrow\ket{\xx}$ which is valid assuming the detuning between them is much larger than the cavity-exciton coupling strength, which is the case for this system. The system Hamiltonian, $H_{\mathrm{sys}}$, under the rotating wave and dipole approximations  \cite{yamamoto,mandelandwolf}  is then
\begin{widetext}
\begin{eqnarray}\label{eqn:Hamiltonian}
H_{\mathrm{sys}}&=&\omega_0 \op{\bx}{\bx}+\omega_2\op{\xx}{\xx}+\omega_4\op{\xy}{\xy}+\omega_A(\nxa+\nya)+\omega_B(\nxb+\nyb)\nonumber\\
&& {}+\frac{i}{2} (q_1\op{\xx}{\bx}\Ax^\dagger+q_2\op{\g}{\xx}\Bx^\dagger+
q_3\op{\xy}{\bx}\Ay^\dagger+q_4\op{\g}{\xy}\Ax^\dagger-\hc),
\end{eqnarray}
\end{widetext}
where $a_j$ and $\n_j=a_j^\dagger a_j$ are respectively the photon annihilation operator and photon number operator for mode $j$, and for convenience we take $\hbar=1$.  We transform to  an interaction picture defined by $H_0=\frac{\omega_0}{2}\excitationnumber$, where 
\begin{eqnarray}
\excitationnumber&=&2\op{\bx}{\bx}+\op{\xx}{\xx}+\op{\xy}{\xy}\nonumber\\
&&{}+\nxa+\nya+\nxb+\nyb\label{eqn:excitationnumber}
\end{eqnarray}
is the number of excitations in the system.  The interaction Hamiltonian, $H=e^{i H_0 t}H_{\mathrm{sys}}e^{-i H_0 t}-H_0$, is given by 
\begin{widetext}
\begin{eqnarray}\label{eqn:RotatingHamiltonian}
H&=&{}-\xi\op{\xx}{\xx}-(\xi+\Delta)\op{\xy}{\xy}+(\xi+\delta_A)(\nxa+\nya)-
(\xi+\delta_B)(\nxb+\nyb)\nonumber\\
&&{}+\frac{i}{2} (q_1\op{\xx}{\bx}\Ax^\dagger+q_2\op{\g}{\xx}\Bx^\dagger+
q_3\op{\xy}{\bx}\Ay^\dagger+q_4\op{\g}{\xy}\Ax^\dagger-\hc),
\end{eqnarray}
\end{widetext}
where $2\xi=\omega_1-\omega_2$ is the biexciton shift, $\Delta=\omega_3-\omega_1=\omega_2-\omega_4$ is the doublet splitting due to dot asymmetry, $\delta_A=\omega_A-\omega_1$ is the detuning between cavity mode $A$ and transition frequency $\omega_1$ and $\delta_B=\omega_2-\omega_B$ is the detuning between transition frequency $\omega_2$ and cavity mode $B$.  These frequencies are shown schematically in \fig{fig:Spectrum}..

We now define a ``balanced cavity" to be one for which  the two cavity modes fall directly in between each of the doublets ($\delta_A=\delta_B=\Delta/2$) and the exciton-cavity coupling constants are matched ($q_1=q_3$ and $q_2=q_4$).  An ``unbalanced cavity" is one for which $\delta_{A,B}\neq\Delta/2$, and ``unbalanced coupling"   means that $q_1\neq q_3$ or $q_2\neq q_4$).  We will show later that a balanced cavity accomplishes the required ``which-path" erasure.

\begin{figure*}
\includegraphics{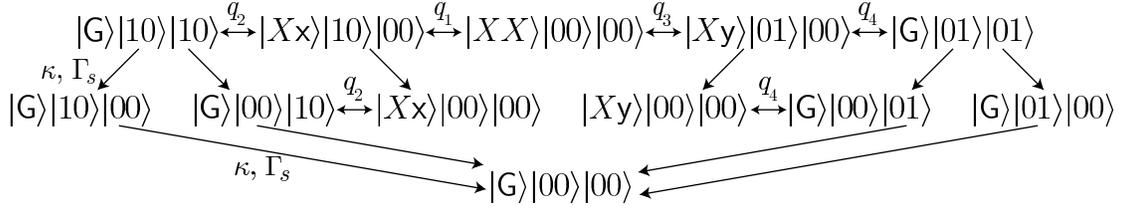}
\caption{\label{fig:EmissionProcess} The basis $\basis$ for the evolution of the system along with transitions generated by the Hamiltonian $H$ and cavity leakage.  Coupling strengths between states are indicated.  The top line of states spans the subspace of two excitations (i.e. exciton number plus  photon number), the middle line spans the subspace of one excitation, and the single state on the last line spans the subspace of zero excitations.}
\end{figure*}

The dynamics of states under the action of the time evolution operator, $e^{-i H t}$, generated by the Hamiltonian $H$ is closed in the 12-dimensional space spanned by the basis $\basis$, which is shown in \fig{fig:EmissionProcess}.

Finally, we assume that the initial state of the system is biexcitonic, $\ket{\psi(0)}=\basisket{\bx}{00}{00}$.

\section{Derivation of master equation} 

The theory of open quantum systems has been well studied (see e.g. \cite{gar00,wis94}), and we adopt this formalism to analyse the exciton-cavity system interacting with the external continuum modes and measurement devices outside the cavity.  

Wiseman \cite{wis94} gives an expression for the master equation for the conditional density matrix, $\rho_c$, for a single measurement channel using imperfect detectors, while Gardiner and Zoller \cite{gar00} give a similar expression for many channels with perfect detection on each channel.  Generalizing these results to an $n$ channel conditional master equation with arbitrary efficiency detectors on each channel results in the conditional master equation
\begin{widetext}
\begin{equation}
\label{eqn:NChannelCME}
d {\rho}_c=- i \commute{H}{\rho_c} d t +\sum_{j=1}^{n}\left\{\left(\eta_j \tr{\jump_j \rho_c}\rho_c 
+(1-\eta_j)\jump_j \rho_c -\frac{1}{2}(c_j^\dagger c_j \rho_c+ \rho_c c_j^\dagger c_j)\right)d t
+\left(\frac{\jump_j \rho_c}{\tr{\jump_j \rho_c}}-\rho_c\right) d N_j\right\},
\end{equation}
\end{widetext}
where $H$ is the interaction Hamiltonian for the system, $c_j$ is the system operator through which the system couples to channel $j$, $\jump_j \rho_c\equiv c_j \rho_c c_j^\dagger$ is the jump operator for channel $j$, $\eta_j$ is the detection efficiency of jump processes on channel $j$, and $d N_j$ is the jump increment.  For the case where $\eta_j=1$ for all $j$, this equation reproduces the result in Gardiner and Zoller  \cite[\S 11.3.8.d]{gar00}, and for $n=1$ it reproduces the result of Wiseman \cite[\S 4.1.2]{wis94}.

For the biexciton decay, there are several baths with which the exciton-cavity system is coupled.  Firstly, the cavity modes decay at a rate $\kappa$ in order to couple the photons generated in the emission process to the outside world.  This decay mode is coupled to the four cavity modes, where the system coupling operators are $c_1=\sqrt{\kappa}\Ax, c_2=\sqrt{\kappa}\Ay, c_3=\sqrt{\kappa}\Bx$ and $c_4=\sqrt{\kappa}\By$.  To quantify the effect of  the cavity in erasing the frequency information, in what will follow, these channels are assumed to be perfectly detected, $\eta_1=\dots=\eta_4=1$.

Secondly, there may be spontaneous emission into photon modes apart from those of the cavity.  This decay channel couples via similar system operators, but with different decay rates, so that $c_5=\sqrt{\Gamma_s}\ket{\xx}\bra{\bx},c_6=\sqrt{\Gamma_s}\ket{\xy}\bra{\bx}, c_7=\sqrt{\Gamma_s}\ket{\g}\bra{\xx}$ and $ c_8=\sqrt{\Gamma_s}\ket{\g}\bra{\xy}$.  These channels are considered to be inaccessible to an observer, so we set the detection efficiency to zero, $\eta_4=\dots=\eta_8=0$.  For later sections, we will refer to these channels as ``leakage channels".  

Finally, we will add a phenomenological dephasing acting on the two exciton states $\ket{\xx}$ and $\ket{\xy}$.  This is to simulate the effect of some unspecified bath (e.g. phonons) that is able to distinguish the intermediate excitonic state during the decay process.  The system operators to which this bath couples are assumed to be $c_9=\sqrt{\Gamma_d}\op{\xx}{\xx},c_{10}=\sqrt{\Gamma_d}\op{\xy}{\xy}$.  Again, these channels are inaccessible to observer, so the detection efficiency is zero, $\eta_9=\eta_{10}=0$.

Since $\eta_j=0$ for channels 5 through 10, from \cite{wis94} we have $E[d N_j(t)]=\eta_j \tr{\jump_j \rho_c(t)} dt=0$ for $j=5\dots10$, and since $d N_j(t)$ is non-negative, $d N_j(t)=0$ for $j=5\dots10$.  In accordance with the assumptions regarding channel efficiencies made above, the conditional master equation between photon detections (i.e. $d N_j=0, j=1\dots4$), becomes
\begin{eqnarray}
\label{eqn:CME}
\dot{\rho}_c=&-&i \commute{H}{\rho_c}+\sum_{j=1}^{4}\tr{\jump_j \rho_c}\rho_c\nonumber
\\&+&\sum_{j=5}^{10}\jump_j \rho_c -\sum_{j=1}^{10}\frac{1}{2}(c_j^\dagger c_j \rho_c+ \rho_c c_j^\dagger c_j).
\end{eqnarray}
The second term in this equation is non-linear in $\rho_c$, reflecting the fact that the evolution is conditional on the system not emitting a photon.  For computational purposes, we convert \eqn{eqn:CME} into an equivalent linear equation for an unnormalized density matrix $\tilde\rho$ by defining $\rho_c(t)=f(t)\tilde\rho(t)$, where $f(t)$ is a scalar function to be determined.  Substituting this into \eqn{eqn:CME} gives
\begin{eqnarray}
f \dot{\tilde{\rho}}+\dot{f} \tilde\rho=&-&i f \commute{H}{\tilde\rho}+f^2\sum_{j=1}^{4}\tr{\jump_j \tilde\rho}\tilde\rho\nonumber
\\&+&f\sum_{j=5}^{10}\jump_j \tilde\rho -f \sum_{j=1}^{10}\frac{1}{2}(c_j^\dagger c_j \tilde\rho+ \tilde\rho c_j^\dagger c_j).
\end{eqnarray}
Collecting terms that are proportional to $f$ and requiring that others vanish gives the linear, unnormalized semi-conditional (i.e. conditioned on only a subset, $j=1\dots4$, of the channels) master equation
\begin{equation}
\label{eqn:linearCME}
\dot{\tilde\rho}=-i \commute{H}{\tilde\rho}
+\sum_{j=5}^{10}\jump_j \tilde\rho - \sum_{j=1}^{10}\frac{1}{2}(c_j^\dagger c_j \tilde\rho+ \tilde\rho c_j^\dagger c_j),
\end{equation}
along with the constraint equation for $f$
\begin{equation}\label{eqn:feqn}
\dot{f} \tilde\rho=f^2\sum_{j=1}^{4}\tr{\jump_j \tilde\rho}\tilde\rho.
\end{equation}
This can be integrated to give $f=-(\int^t dt \sum_{j=1}^{4}\tr{\jump_j \tilde\rho})^{-1}.$
Taking the trace of \eqn{eqn:linearCME} gives 
$\tr{\dot{\tilde\rho}}=-\sum_{j=1}^{4}\tr{\jump_j \tilde\rho}$, and so we see that $f=\tr{\tilde\rho}^{-1}$, which is just the normalization condition for $\rho_c$, i.e. $\rho_c=\tilde\rho/\tr{\tilde\rho}$ as required.

\section{Quantifying two-photon entanglement}

We now develop a measure of the performance of the cavity in erasing ``which-path" information.  A polarisation entangled photon pair is an archetypal example of a two-qubit  system.  Such bipartite systems have been studied extensively \cite{nie00}, and in particular, the entanglement of such pure bipartite system is well quantified by the von Neumann entropy of one subsystem.  

\begin{figure}
\includegraphics[width=8cm]{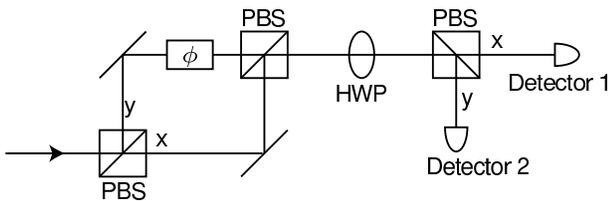}
\caption{\label{TwoPhotonInterferometer} Schematic of a polarisation sensitive interferometer.  A polarising beam splitter (PBS) splits the beam into $\x$- and $\y$-polarised paths and a relative phase $\phi$ is added to one path.  The paths are recombined, and the polarisations are rotated by $\pi/4$ using a half-wave plate (HWP), then detected with polarisation-sensitive single-photon detectors}
\end{figure}

In the system we are concerned with, there is some subtlety however, since although the system-plus-continuum evolves to a two photon state, this is only determined once a measurement has been performed, observing both photons.    Before the measurement, the system evolves in a much larger Hilbert space, so it is not entirely trivial to adapt measures like entropy to the case of interest here, not least because under certain circumstances the photon-pair is described by a mixed state.  Instead we use an operational measure -- the visibility.  This arises naturally by considering the result of two-photon coincidence counting at the output of a polarisation sensitive interferometer, depicted in \fig{TwoPhotonInterferometer}, through which the photon-pair is directed.  

It is straightforward to show that a pure, entangled state of the form $\ket{\x\x}+\ket{\y\y}$ passing through such an interferometer with a $\phi$ phase-shift (per photon) on one arm will exhibit interference fringes in the two-photon coincidence counts between the detectors, with the coincidence count rate proportional to $1-\cos(2\phi)$.  The factor of 2 in the argument of the cosine is a direct manifestation of the two particle nature of the state, and this has been observed experimentally \cite{eda02}.  

Conversely, neither a completely mixed state such as $\op{\x\x}{\x\x} +\op{\y\y}{\y\y}$ nor a pure, non-entangled state like $\ket{\x\x}$ will display interference fringes as $\phi$ is varied.  It is intuitively clear from these two examples that the visibility of the interference fringes is an operational measure of both the purity and entanglement of the input two-photon state, and is dependent on the off-diagonal elements of the density matrix, which are zero for non-entangled or completely mixed states.

Interferometric methods for estimating entanglement have been discussed by Ekert and Horodecki \cite{eke01}.  They argue that $d^2-1$ separate types of interferometric experiment are required to estimate the entanglement of a pair of particles, $d$ being the dimension of the Hilbert space for each particle.  For the case of interest to us $d=2$, so we expect three parameters will be sufficient to place bounds on the entanglement of the photon pair.  In fact, we assume that anti-correlated states such as $\ket{\x\y}$ are never produced, which is a roughly consistent with  experimental observations showing that inter-exciton transitions are rare, \cite{ste02}, and so the number of experiments required is reduced to one.  That is, we only need to measure a single visibility fringe in order to quantify the two-photon entanglement.

\subsection{Interferometry}
Quantitatively, we relate the output continuum field annihilation operators of the half wave plate, $b_{\x,\omega}', b_{\y,\omega}'$,  to the interferometer input field operators, $b_{\x,\omega}, b_{\y,\omega}$, according to
\begin{equation}\label{eqn:InterferometerTransformation}
\begin{bmatrix}
b_{\x,\omega}'\\
b_{\y ,\omega}'
\end{bmatrix}=
\frac{1}{\sqrt{2}}\begin{bmatrix}
e^{i \phi} & 1\\
-1 & e^{-i \phi}
\end{bmatrix}\cdot
\begin{bmatrix}
b_{\x ,\omega}\\ b_{\y ,\omega}
\end{bmatrix}.
\end{equation}
We adopt the notation that a prime on an operator indicates that it is transformed consistently with  \eqn{eqn:InterferometerTransformation}, for example $\n_{\x,\omega}'=(b_{\x,\omega}')^\dagger b_{\x,\omega}'$.

The expectation of a two-photon coincidence measurement by detectors 1 and 2 will in general be given by the normally ordered (denoted by :\dots:) two-time correlation function $\langle : \n_{\poli,\omega_B}'(t_B) \n_{\polj,\omega_A}'(t_A) :\rangle_c$, where $\poli, \polj\in\{\x,\y\}$ \cite{gar00, wis93, wis94}.  The subscript $c$ denotes the fact that the expectation is conditioned on the system having emitted zero photons in the interval $[0,\min\{t_A,t_B\})$, and the ordering of the operators in the correlation function will depend on the ordering of $t_A$ and $t_B$.

We may relate the cavity field output operators for the continuum mode $l$ to the cavity input operators and the internal cavity operator according to $b_\out(l,t)-b_\inop(l,t)=\sqrt{\kappa} a_l(t)$.  We will also assume that the cavity input is the vacuum so $\langle b_{\inop}^\dagger(l,t)b_{\inop}(l,t')\rangle=0$ \cite{gar00}.  Thus normally ordered expectations of continuum modes may be replaced by normal-and-time ordered expectations of internal cavity modes, multiplied by a  suitable power of $\sqrt{\kappa}$.  More detailed discussion of this point is given in Gardiner \& Zoller \cite{gar00}.

For example, the conditional expectation of detecting consecutive photons at detector 1 will be given by $\langle :\n_{\x,\omega_B}'(t_B) \n_{\x,\omega_A}'(t_A):\rangle_c$, and if $t_B>t_A$, then 
\begin{eqnarray}\label{eqn:TwoPhotonCorrelator}
&&\langle :\n_{\x,\omega_B}'(t_B)\n_{\x,\omega_A}'(t_A):\rangle_c\nonumber\\
&&{}=\tr{\jump_{\x,\omega_B}'(t_B)\T(t_B,t_A)\{\jump_{\x,\omega_A}'(t_A)\rho_c(t_A)\}},\\
&&{}=\kappa^2 \tr{(\Bx')^\dagger  \Bx'  \T(t_B,t_A)\{\Ax'\rho_c(t_A) (\Ax')^\dagger\}},\nonumber
\end{eqnarray}
where $\T(t_B,t_A)$ is the time evolution operator, which evolves the system from time $t_A$ to time $t_B$, and for open systems it is non-unitary \cite{gar00}.  Very similar expressions may be derived for the case where $t_B<t_A$.

\subsection{Visibility}
Since the transformed operators in \eqn{eqn:TwoPhotonCorrelator} depend on $\phi$ according to \eqn{eqn:InterferometerTransformation}, we see that the quantity $\langle :\n_{\x,\omega_B}'(t_B) \n_{\x,\omega_A}'(t):\rangle$ must also depend on $\phi$.  Many of the cross terms vanish, leaving the result
\begin{equation}\label{eqn:ExpectationValue}
\langle :\n_{\x,\omega_B}'(t_B)\n_{\x,\omega_A}'(t_A):\rangle =\kappa^2(x+y+e^{2 i \phi} z+ e^{-2 i \phi} z^*),
\end{equation}
where 
\begin{widetext}
\begin{eqnarray}\label{eqn:xyz}
x&\equiv&\braket{:\Bx^\dagger\Bx\Ax^\dagger\Ax:}=\tr{\Bx^\dagger\Bx\T\{\Ax\rho_c\Ax^\dagger\}}\in[0,1]\nonumber\\ 
y&\equiv&\braket{:\By^\dagger\By\Ay^\dagger\Ay:}=\tr{\By^\dagger\By\T\{\Ay\rho_c\Ay^\dagger\}}\in[0,1]\nonumber\\ 
z&\equiv&\braket{:\By^\dagger\Bx\Ay^\dagger\Ax:}=\tr{\By^\dagger\Bx\T\{\Ax\rho_c\Ay^\dagger\}}\in \mathbb{C},
\end{eqnarray}
\end{widetext}
which all depend on $t_A$ and $t_B$ though this notation has been dropped for brevity.  We can define the visibility, $\vis$, from this expression to be the amplitude of the interference fringes divided by the mean (averaged over $\phi$) and it is
\begin{equation}
\vis(t_A,t_B)=\frac{2|z|}{x+y}.\label{eqn:Visibility}
\end{equation}
Conceptually, $\vis$ is the visibility of fringes generated by post-selecting photon pairs that arrive within the two-time window  $(t_A,t_A+dt)\times(t_B,t_B+dt)$ as $\phi$ varies.  We note that we may compute the visibility directly from $\tilde\rho$ by making the definition $\tilde x\equiv\tr{\Bx^\dagger\Bx\T\{\Ax\tilde\rho\Ax^\dagger\}}$, with similar definitions for $\tilde y$ and $\tilde z$, so that an equivalent expression for the visibility is
\begin{equation}\label{eqn:Visability}
\vis(t_A,t_B)=\frac{2|\tilde z|}{\tilde x+\tilde y}.
\end{equation}

\subsection{Probability density}
We may also compute the joint probability density, $\prob$, for detecting a photon pair within the two-time window $(t_A,t_A+dt)\times (t_B,t_B+dt)$, as given in \cite[\S 11.3.7 (d)]{gar00},
\begin{eqnarray}
&&\prob(t_A,t_B)=\left(\sum_{\poli,\polj} \langle :\n_{\poli,\omega_B}'(t_B) \n_{\polj,\omega_A}'(t_A):\rangle_c\right)\nonumber\\
&&\hphantom{\prob(t_A,t_B)=}{}\times\left(1- \int_0^{t_A} d\tau\sum_j\jump_j'\T(\tau,0){\rho(0)}\right), \nonumber\\
&&= \sum_{\poli,\polj} \langle :\n_{\poli,\omega_B}'(t_B) \n_{\polj,\omega_A}'(t_A):\rangle_c \tr{\tilde\rho(t_A)}, \nonumber\\
&&=\kappa^2 \sum_{\poli,\polj} \tr{\jump_{\poli,\omega_B}'(t_B)\T(t_B,t_A)\{\jump_{\polj,\omega_A}'(t_A)\tilde\rho(t_A)\}}\nonumber,\\
&&=\kappa^2(\tilde x+\tilde y)\label{eqn:Probability}
\end{eqnarray}
where we have again assumed $t_A<t_B$, although similar expressions may easily be derived for $t_A>t_B$.  The first factor in in the first equality is just the conditional probability density for either detector to register at times $t_A$ and $t_B$ given no emission beforehand, and the second factor is the probability of emitting zero photons in the interval $[0,t_A)$.  The second equality follows from \eqn{eqn:feqn} and later. The third equality follows from \eqn{eqn:TwoPhotonCorrelator} and recalling the fact that $\rho_c(t)=\tilde\rho(t)/\tr{\tilde\rho(t)}$.  Finally, \eqn{eqn:Probability} shows that the probability density does not depend on $\phi$ -- detecting a photon pair after the interferometer occurs with the same probability density as detecting a photon pair before the interferometer, as expected.

We also define the quantity 
\begin{equation}
P=\int_{0}^{\infty} \int_{0}^{\infty}\prob(t_{A},t_{B})dt_{A}dt_{B}.
\end{equation}
In the presence of spontaneous emission into non-cavity photon modes, $P<1$, indicating that not all biexciton decay events will be detected by the photodetectors following the interferometer.  We therefore interpret $P$ as the reduction factor of the two-photon detection rate, as compared with the biexciton  pumping rate.

\subsection{Mean visibility}

We now define the mean visibility, which is a figure of merit for the degree of entanglement between the photon pair,
\begin{eqnarray}\label{eqn:MeanVisibility}
\meanvis&=&\frac{1}{P}\int_{0}^{\infty} \int_{0}^{\infty} \vis(t_{A},t_{B})\prob(t_{A},t_{B}) dt_{A}dt_{B},\nonumber\\
&=&\frac{2\kappa^2}{P}\int_{0}^{\infty} \int_{0}^{\infty}|\tilde z(t_{A},t_{B})|dt_{A}dt_{B},
\end{eqnarray}
where we have divided by $P$ so as to only count those decay events that are detected through the interferometer.
If the visibility is unity (i.e. perfect erasure of ``which-path'' information), then $\meanvis=1$, since the probability density is normalised by $P$.  On the other hand, if $\vis$ is less than unity, then $\meanvis$ will be also, so performing a two-photon interference experiment with all photon pairs produced will result in fringes of visibility $\meanvis<1$.

We see from \eqns{eqn:Visibility}, (\ref{eqn:Probability}) and (\ref{eqn:MeanVisibility}) that the quantities we are interested in may all be determined directly from $\tilde \rho$, which makes calculations we perform in following sections simpler.

\subsection{Phase accumulation}

Whilst the visibility is a very important measure of the success of the scheme, since the initial state of the system, $\ket{\xx}\ket{00}\ket{00}$, is not an energy eigenstate, during the emission process, phase will accumulate at  different rates on the $\x\x$- and $\y\y$-decay paths.  The phase difference accumulated between each decay path depends on the emission times of the two photons and is given by $\varphi\equiv\arg\{z(t_A,t_B)\}=\arg\{\tilde z(t_A,t_B)\}$, 
corresponding to emission of a state of the form $\ket{\x\x}+e^{i \varphi(t_A,t_B)}\ket{\y\y}$.  Since, for a given apparatus, $\varphi$ depends only on the emission times $t_A$ and $t_B$, this may be calibrated or computed, and hence accounted for, before an interference experiment (or whatever else is intended for the output photon pair) is done.  If this phase is ignored, then the mean visibility will be lower than $\meanvis$ since the description of the phase-averaged state will be mixed.

\subsection{Relation to Bell-inequality violations}

\begin{figure}
\includegraphics{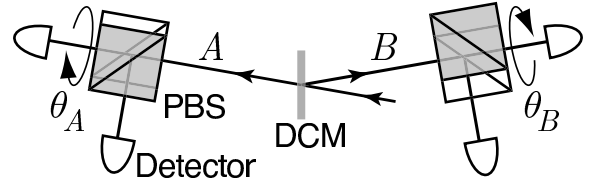}
\caption{
\label{fig:CHSH}  Schematic setup for performing a CHSH Bell-inequality measurement.  Photons of frequency $\omega_A$ and $\omega_B$ are split with a dichroic mirror (DCM) then travel along paths $A$ and $B$ respectively, and are measured by rotated polarisation-sensitive detectors.}
\end{figure}

Instead of passing the photon pair through an interferometer, we could imagine using an ensemble of such states to measure Bell-inequality violations.  In particular, we consider violations of a Clauser--Horne--Shimony--Holt (CHSH) inequality, where each photon is measured in one of two non-orthogonal bases specified by the angles $\theta_A$ and $\theta_A'$ for the photon at of frequency $\omega_A$ and $\theta_B$ and $\theta_B'$ for the photon at of frequency $\omega_B$ \cite{wal94}, as depicted in \fig{fig:CHSH}.

In terms of mode operators, the CHSH inequality requires the knowledge of correlation functions of the form
\begin{equation}\label{eqn:CHSH_expectation}
E(\theta_A,\theta_B)=\frac{
\braket{:(d_+^\dagger d_+ -d_-^\dagger d_-)(c_+^\dagger c_+ -c_-^\dagger c_-):}}
{\braket{:(d_+^\dagger d_+ +d_-^\dagger d_-)(c_+^\dagger c_+ +c_-^\dagger c_-):}},
\end{equation}
where photon mode annihilation operators $c$ and $d$ are defined as
\begin{eqnarray}
c_+&=&\sin(\theta_A) \Ay+\cos(\theta_A) \Ax,\nonumber\\
c_-&=&\cos(\theta_A) \Ay-\sin(\theta_A) \Ax,\nonumber\\
d_+&=&\sin(\theta_B) \By+\cos(\theta_B) \Bx,\nonumber\\
d_-&=&\cos(\theta_B) \By-\sin(\theta_B) \Bx.\nonumber
\end{eqnarray}
We have not explicitly included times in these expressions, but we note that operators $\Ai$ act at time $t_A$ and $\Bi$ at $t_B$.  It is straightforward to show that
\begin{eqnarray}
c_+^\dagger c_+ +c_-^\dagger c_-&=&\Ay^\dagger\Ay+\Ax^\dagger\Ax,\nonumber\\
c_+^\dagger c_+ -c_-^\dagger c_-&=&\cos(2\theta_A)(\Ay^\dagger\Ay+\Ax^\dagger\Ax)\nonumber\\&&{}+\sin(2\theta_A)(\Ay^\dagger\Ax+\Ax^\dagger\Ay),\nonumber
\end{eqnarray}
with similar results for $d_\pm$.  As mentioned earlier, many cross-terms in the numerator and denominator of \eqn{eqn:CHSH_expectation} cancel for the physical situation we consider, e.g. \mbox{$\braket{:\Ax^\dagger\Ax\By^\dagger\By:}=0$}, so we can write it as
\setlength{\arraycolsep}{0pt}
\begin{eqnarray}
&&E(\theta_A,\theta_B)\nonumber\\&&=\cos(2\theta_A)\cos(2\theta_B)
+\frac{z+z^*}{x+y}\sin(2\theta_A)\sin(2\theta_B),\nonumber\\
&&=\cos(2\theta_A)\cos(2\theta_B)+\vis\cos(\varphi)\sin(2\theta_A)\sin(2\theta_B), \label{eqn:EV}
\end{eqnarray}
\setlength{\arraycolsep}{2pt}
where $x, y$ and $z$ are defined in \eqn{eqn:xyz}.

\begin{figure}
\includegraphics[width=8.5cm]{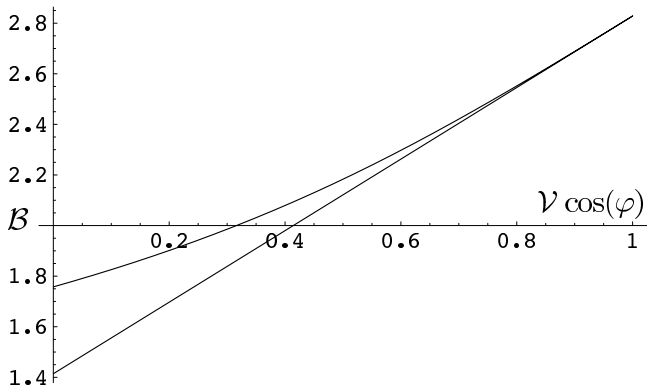}
\caption{
\label{fig:BellViolation}  The quantity $\Bell$ versus $\vis$.  The lower curve is $\Bell$ for $\vartheta=\pi/8$.  The upper curve is the maximum value of $\Bell$ for the corresponding value of $\vis$, allowing $\vartheta$ to change with $\vis$.}
\end{figure}
We now define the quantity
\begin{equation}\label{eqn:BellE}
\Bell\equiv E(\theta_A,\theta_B) -E(\theta_A,\theta_B') +E(\theta_A',\theta_B') +E(\theta_A',\theta_B),
\end{equation}
which, for classically correlated states satisfies $\Bell\leq2$ \cite{wal94}.  This inequality is violated by certain entangled states such as Bell states, which are emitted from the biexciton system.  We note that $\Bell$ depends on $t_A$ and $t_B$, since $\vis$ and $\varphi$ does, but we leave out the explicit notation. From \eqns{eqn:EV} and (\ref{eqn:BellE}) we derive a  linear relationship between $\Bell$ and $\vis$ given by
\begin{eqnarray}
\Bell&=&\cos (2{{\theta }_A})
   \left( \cos (2{{\theta }_B}) - 
     \cos (2\theta _B') \right) \nonumber\\&&{} + 
  \cos (2{{\theta }_A'})
   \left( \cos (2{{\theta }_B}) + 
     \cos (2{\theta }_B' )\right) \nonumber\\&&{}+ 
\vis\cos(\varphi)\big\{ \sin (2{{\theta }_A})
      \left( \sin (2{{\theta }_B}) - 
        \sin (2{{\theta }_B'}) \right)   \nonumber\\&&\hphantom{+ 
\{}{} + 
     \sin (2{{\theta }_A'})
      \left( \sin (2{{\theta }_B}) + 
        \sin (2{{\theta }_B'}) \right)  \big\}  
\end{eqnarray}

We consider a special choice of angles that maximally violate the CHSH inequality, $\theta_A=0, \theta_B=\vartheta, \theta_A'=2\vartheta$ and $\theta_B'=3\vartheta$, and using \eqn{eqn:EV} in the expression for $\Bell$ we find that
\begin{eqnarray}\label{eqn:Bell}
\Bell&=&\cos(2\vartheta )\left( 3 - 2\cos (4\vartheta ) + 
     \cos (8\vartheta ) \right)  \nonumber\\&&{}+ 
  8{\cos (2\vartheta )}^3{\sin (2\vartheta )}^2 \vis \cos(\varphi).
\end{eqnarray}
For $\vis=1$, this gives $\Bell=3\cos(2\vartheta)-\cos(6\vartheta)$, which has a maximum at $\vartheta=\pi/8$ of $2\sqrt{2}>2$, violating the CHSH inequality.  At $\vartheta=\pi/8$ \eqn{eqn:Bell} reduces to 
\begin{equation}\Bell=\sqrt{2}(1+\vis \cos(\varphi))\label{eqn:simpleBell}
\end{equation}
which is plotted  in \fig{fig:BellViolation}  (lower curve). We also show  the maximum value of $\Bell$ for each value of $\vis$ (upper curve), allowing $\vartheta$ to vary.  The upper curve crosses $\Bell=2$ at $\vis\cos(\varphi)\approx0.316$, whilst the lower curve crosses at $\vis\cos(\varphi)=\sqrt{2}-1\approx0.414$.  

From these results we see that $\vis$ and $\Bell$ are very closely related quantities.  Since $\Bell$ may be computed from $\vis$ for arbitrary angles, we will base our computations on $\vis$, from which a reasonable estimate for the maximum value $\Bell$ may be evaluated using \eqn{eqn:simpleBell}.  Finally, we define the quantity $\meanBell$ as 
\begin{equation}
\meanBell=\frac{1}{P}\int_{0}^{\infty} \int_{0}^{\infty} \Bell(t_{A},t_{B})\prob(t_{A},t_{B}) dt_{A}dt_{B},
\end{equation}
where we have again divided by $P$ in order to count only those photon pairs that are detected in the experiment.  

\subsection{Phase-averaged Bell-inequality violation}

Comparing  $\meanBell$ with \eqn{eqn:Bell} or \eqn{eqn:simpleBell}, we see that we need to compute the quantity
\begin{equation}
\frac{1}{P}\int_{0}^{\infty} \int_{0}^{\infty}\prob \vis \cos(\varphi) dt_{A}dt_{B}.\nonumber
\end{equation}
We generalise this to account for the possibility of adding a fixed relative phase $\phi$ to one decay path (e.g by adding a phase plate on the $\y$-polarised photon path, as in the interferometer stage of \fig{TwoPhotonInterferometer}), so that $\cos(\varphi)\rightarrow\cos(\varphi+\phi)$.  We maximise the above integral over $\phi$ to arrive at the \emph{phase-averaged visibility}
\begin{widetext}
\begin{equation}
\Q=\frac{1}{P}\sqrt{\left(\int_0^{\infty}\int_0^{\infty}\prob\vis\cos(\varphi) dt_A dt_B\right)^2
+\left(\int_0^{\infty}\int_0^{\infty}\prob\vis\sin(\varphi) dt_A dt_B\right)^2}.\label{eqn:Q}
\end{equation}
\end{widetext}

The phase-averaged visibility, $\Q$, gives the visibility of the fringes in a two-photon interference experiment where no attempt is made to resolve the phase accumulation.  In regard to a CHSH-inequality violation experiment without sufficiently fast time-resolved detection, violations may still be seen if $\Q>0.316$, since the functional relationship between $\Q$ and $\meanBell$ is the same as that between $\Bell$ and $\vis cos(\varphi)$, as shown in \fig{fig:BellViolation}.

\section{Analytical result for balanced cavity}
\label{sec:analytic_result}

We now show that for a balanced cavity (i.e. $\delta_A=\delta_B=\Delta/2, q_1=q_3$ and $q_2=q_4$), in the absence of spontaneous emission and dephasing,$\Gamma_d=\Gamma_s=0$, the model predicts that the visibility is unity for all $(t_A,t_B)$.  Since we assume $\Gamma_d=\Gamma_s=0$, we may write the unnormalised density matrix as $\tilde \rho(t)=\ket{\state(t)}\bra{\state(t)}$ and then \eqn{eqn:linearCME} may be written as a Schr\"odinger equation
\begin{equation}\label{eqn:SE}
\frac{d}{dt}\ket{\state(t)}=-i\heff\ket{\state(t)}
\end{equation}
for the state vector, $\ket{\state(t)}$,  with a non-Hermitian effective Hamiltonian given by
\begin{equation}
\heff=H-i \kappa/2(\nxa+\nya+\nxb+\nyb).\label{eqn:heff}
\end{equation}
We write the solution to \eqn{eqn:SE} as $\ket{\state(t)}=e^{-i \heff t}\ket{\state(0)}$.  The smooth evolution is of course punctuated by quantum jumps, corresponding to photon detections following the interferometer.  

Reformulating the equations of motion in terms of quantum trajectories \cite{gar00} has several advantages, and most obviously it reduces the number of unknown quantities, since we can now solve for the state vector rather than the density matrix.  It is straightforward to show that the effective Hamiltonian, $\heff$, only couples states within the same excitation-number subspace.  Coupling between the zero, one and two excitation subspaces (denoted $\subspace{0}, \subspace{1}$ and $\subspace{2}$ respectively) occurs only during the jumps, and the excitation number irreversibly decreases by one at each jump as photons leave the cavity.  Thus, for the smooth evolution between jumps, we may consider the evolution restricted to states within each $\subspace{j}$ independently, and for each $\subspace{j}$ we consider the effective Hamiltonian, $\heff_j$, restricted to that subspace and acting on the state vector $\ket{\state(t)}_j$.

Using the quantum trajectories formalism, we find 
\begin{subequations}
\begin{eqnarray}
\tilde x&=&\inner{\state_\x(t_A, t_B)}{\state_\x(t_A, t_B)}\\
\tilde y&=&\inner{\state_\y(t_A, t_B)}{\state_\y(t_A, t_B)}\\
\tilde z&=&\inner{\state_\y(t_A, t_B)}{\state_\x(t_A, t_B)},
\end{eqnarray}
\end{subequations}
 where, assuming $t_A<t_B=t_A+\tau$ we have defined
\begin{eqnarray}
\ket{\state_\poli(t_A, t_B)}&=&\Bi e^{-i \heff_1 \tau} \Ai \ket{\state(t_A)}\nonumber\\
&=&\Bi e^{-i \heff_1 \tau} \Ai e^{-i \heff_2 t_A}\ket{\state(0)}.\label{eqn:finalstate}
\end{eqnarray}
A very similar expression exists for $t_A>t_B$, and the following reasoning applies equally to both cases.  The state vector $\ket{\state_\poli(t_A, t_B)}\in \subspace{0}$, since the initial condition $\ket{\psi(0)}=\basisket{\bx}{00}{00}\in\subspace{2}$ and the effect of the two annihilation operators in \eqn{eqn:finalstate} is to reduce the excitation number by two.

The one dimensional subspace $\subspace{0}$ is spanned by the system ground state $\basisket{\g}{00}{00}$, so a state $\ket{\state(t)} \in \subspace{0}$ is mapped smoothly to a scalar, $\state(t)$, by the trivial mapping $\state(t)\equiv(\basisbra{\g}{00}{00})\ket{\state(t)}$.  We may therefore write $\tilde x, \tilde y$ and $\tilde z$ in terms of the scalar quantities $\state_\x(t_A, t_B)$ and $\state_\y(t_A, t_B)$: $\tilde x=\state_\x^* \state_\x, \tilde y=\state_\y^* \state_\y$ and $\tilde z=\state_\y^* \state_\x$, where we have dropped the time dependent notation for clarity.

In what follows, we establish that for a balanced cavity, $\state_\x$ and $\state_\y$ are related by a unitary factor.  This means that they have the same amplitude, from which it follows that the visibility is unity for a balanced cavity.  We do this by considering the transformation of the 
effective Hamiltonian and state vector under exchange of the polarisation, $\ket{\xx}\leftrightarrow\ket{\xy}$ and $\ket{01}\leftrightarrow\ket{10}$.  This transformation, denoted hereafter by $^\#$, is just  a permutation on the basis elements, leaving the two elements $\basisket{\bx}{00}{00}$ and $\basisket{\g}{00}{00}$ invariant.  A matrix representation of $^\#$ shows that is both orthogonal and symmetric.

Consider evolution in $\subspace{2}$.  Swapping $\x$ and $\y$ polarisations maps $\heff_2\rightarrow \heff_2^\#=-\heff_2^*$ and $\ket{\state(t)}_2\rightarrow \ket{\state(t)}_2^\#$.  As a result the time evolution operator $(e^{-i \heff_2 t})^\#=e^{-i \heff_2^\# t}=e^{i \heff_2 t}$ when acting on states in $\subspace{2}$.  We note in passing that for an unbalanced cavity/coupling $\heff_2^\#\neq-\heff_2^*$, which is why it is critical that the cavity be balanced this argument to be valid.

A similar result applies to evolution in $\subspace{1}$, except that the effective Hamiltonian $\heff_1$ does not transform under polarisation swapping quite as simply.  Instead, it may be shown that  $\heff_1-H_d\rightarrow \heff_1^\#-H_d^\#=-(\heff_1^*-H_d)$, where $H_d$ is a  Hermitian matrix acting on elements of $\subspace{1}$ and satisfies $[\heff_1,H_d]=0$.   Thus $\heff_1^\#=-\heff_1^*+H_d+H_d^\#$ and $(e^{-i \heff_1 t})^\#=e^{-i \heff_1^\# t}=e^{i \heff_1^* t}e^{-i (H_d+H_d^\#) t}$.  The factor $U_d(t)=e^{-i (H_d+H_d^\#) t}$ is unitary, since $H_d$ is Hermitian.  
In particular, $U_d(t)$ acts on states of the form $\Ai e^{-i \heff t_A}\ket{\state(0)}\in\subspace{1}$ in a simple way: it multiples the state by a time dependent unitary scalar, $e^{i \theta t}$.

Having established the effect of $^\#$ on the time evolution operator acting on $\subspace{1}$ and $\subspace{2}$ we see that for example
\begin{eqnarray}
\ket{\state_\x(t_A, t_B)}^\#&=&(\Bx e^{-i \heff \tau} \Ax e^{-i \heff t_A}\ket{\state(0)})^\#,\nonumber\\
&=&\By e^{-i \heff^\# \tau} \Ay e^{-i \heff^\# t_A}\ket{\state(0)},\nonumber\\
&=&e^{i \theta \tau}\By e^{i \heff^* \tau} \Ay e^{i \heff^* t_A}\ket{\state(0)},\nonumber\\
&=&e^{i \theta \tau}(\By e^{-i \heff \tau} \Ay e^{-i \heff t_A}\ket{\state(0)})^*,\nonumber\\
&=&e^{i \theta \tau}\ket{\state_\y(t_A, t_B)}^*.\label{eqn:psihash}
\end{eqnarray}
This second line follows since $\ket{\state(0)}^\#=\basisket{\bx}{00}{00}^\#=\basisket{\bx}{00}{00}=\ket{\state(0)}$ and $a_\x,\omega^\#=a_\y,\omega$ and the third line follows by considering the arguments in the preceding two paragraphs.

On the other hand, since $\ket{\state_\x(t_A, t_B)}\in\subspace{0}$ it is evident that $\ket{\state_\x(t_A, t_B)}^\#=\ket{\state_\x(t_A, t_B)}$ as $\basisket{\g}{00}{00}$ is invariant under $^\#$.  Together with \eqn{eqn:psihash}, this implies  $\ket{\state_\x(t_A, t_B)}= e^{i \theta \tau}\ket{\state_\y(t_A, t_B)}^*$, and we conclude that $\state_\x=e^{i \theta \tau}\state_\y^* $.  It follows that $\tilde x=\state_\x^*\state_\x=\state_\y^*\state_\y=\tilde y$, and also $|\tilde z|=|\state_\y^*\state_\x|=|e^{-i \theta \tau}\state_\x^2|=\tilde x$.  Using these two results and \eqn{eqn:Visability} we see immediately that for a balanced cavity $\vis=1$, proving that the visibility is unity for all times.

\section{Numerical results}

\begin{figure*}
\subfigure[]{\label{fig:BalancedProb}\includegraphics[width=4cm]{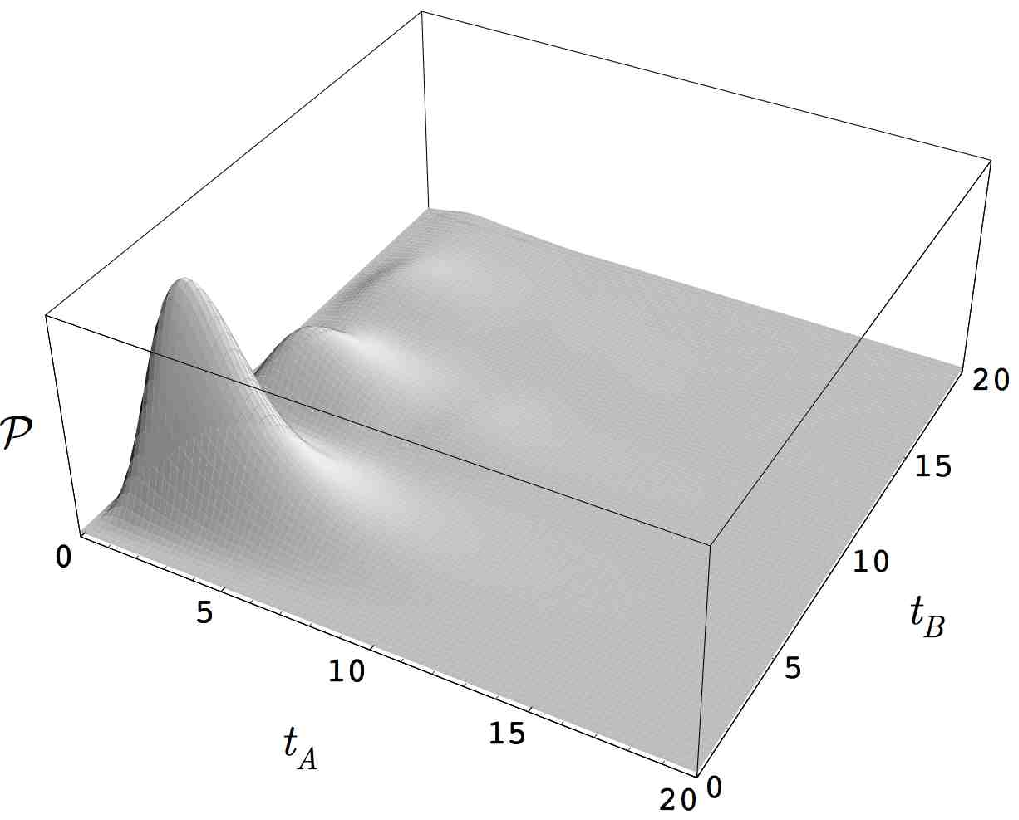}\includegraphics[width=4cm]{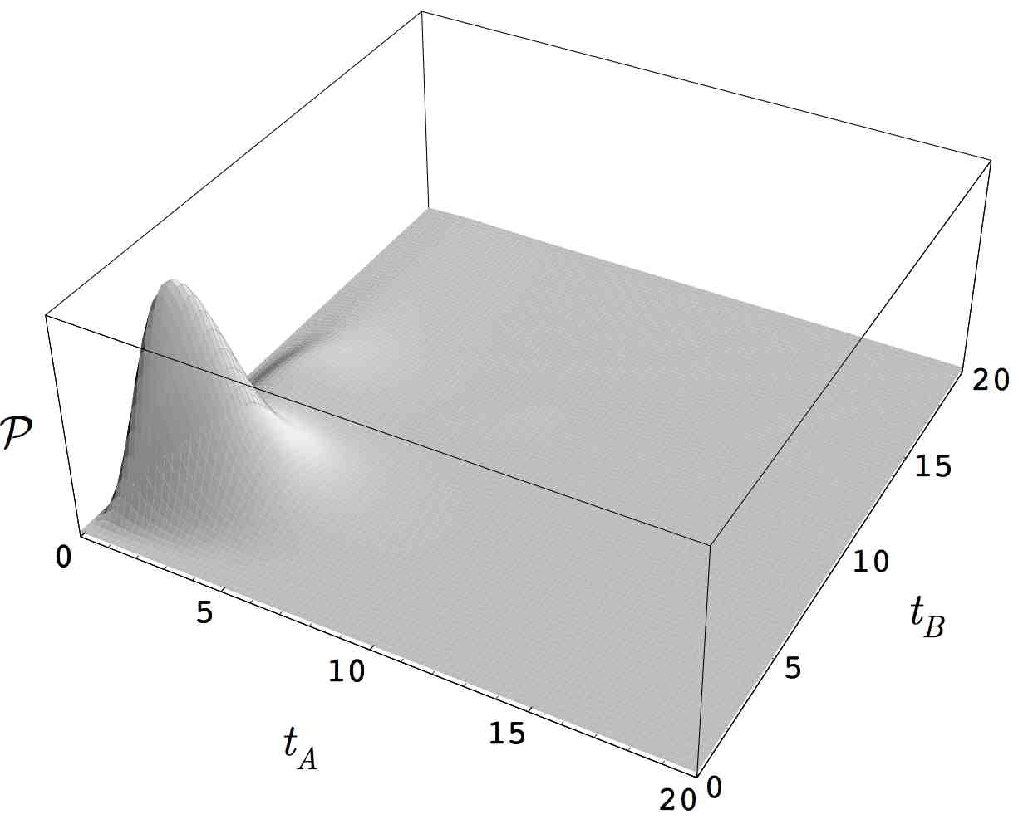}\includegraphics[width=4cm]{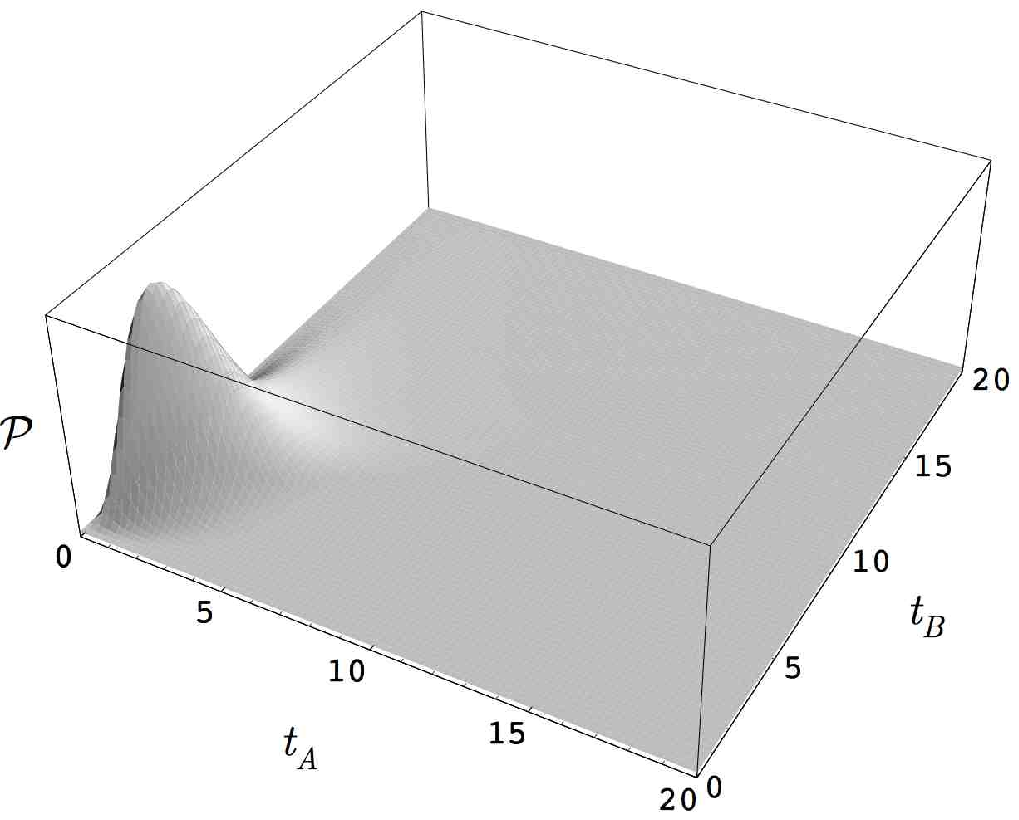}\includegraphics[width=4cm]{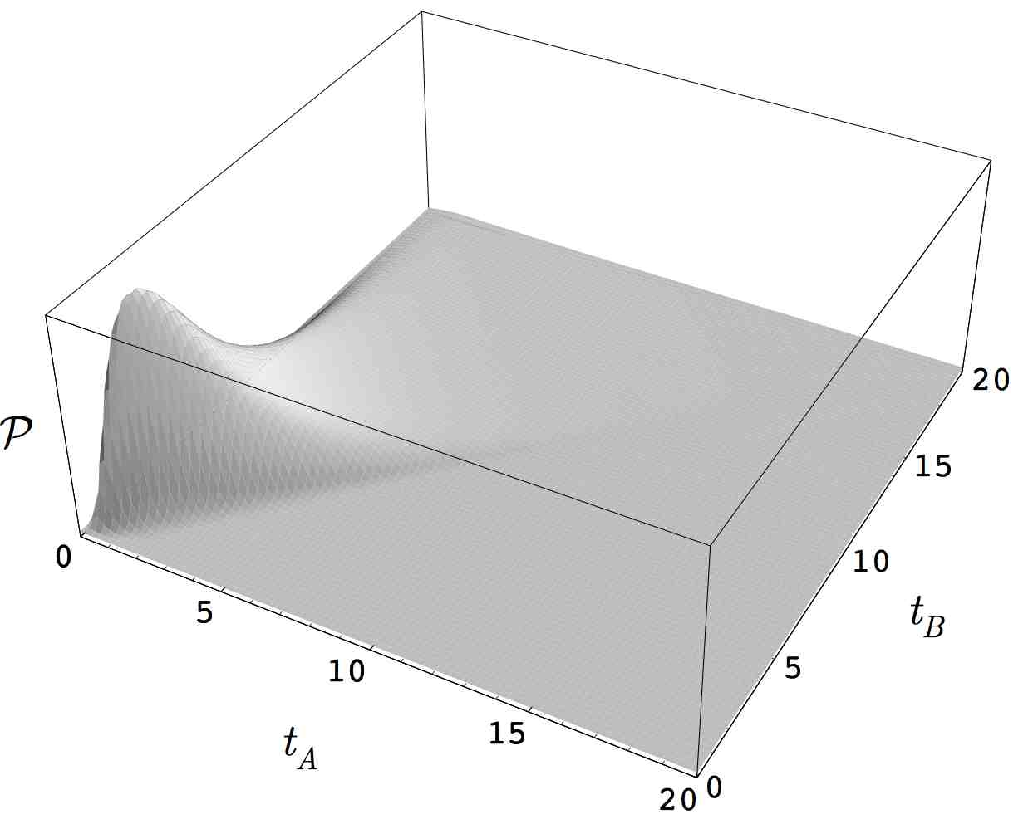}}
\subfigure[]{\label{fig:UnbalancedVis}\includegraphics[width=4cm]{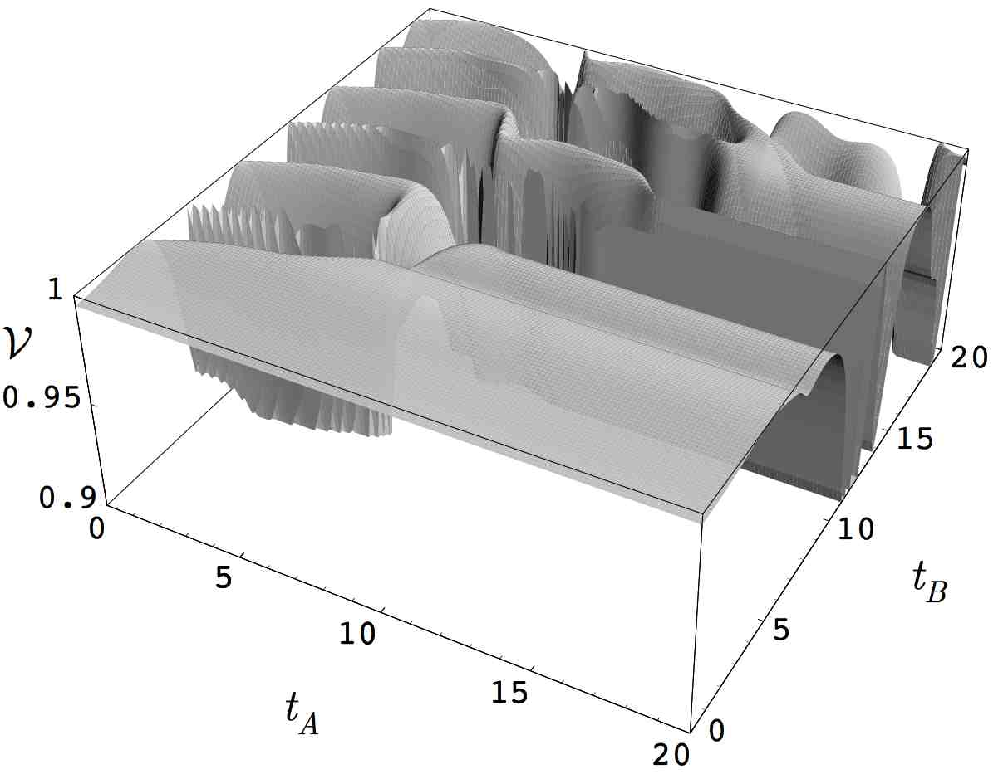}\includegraphics[width=4cm]{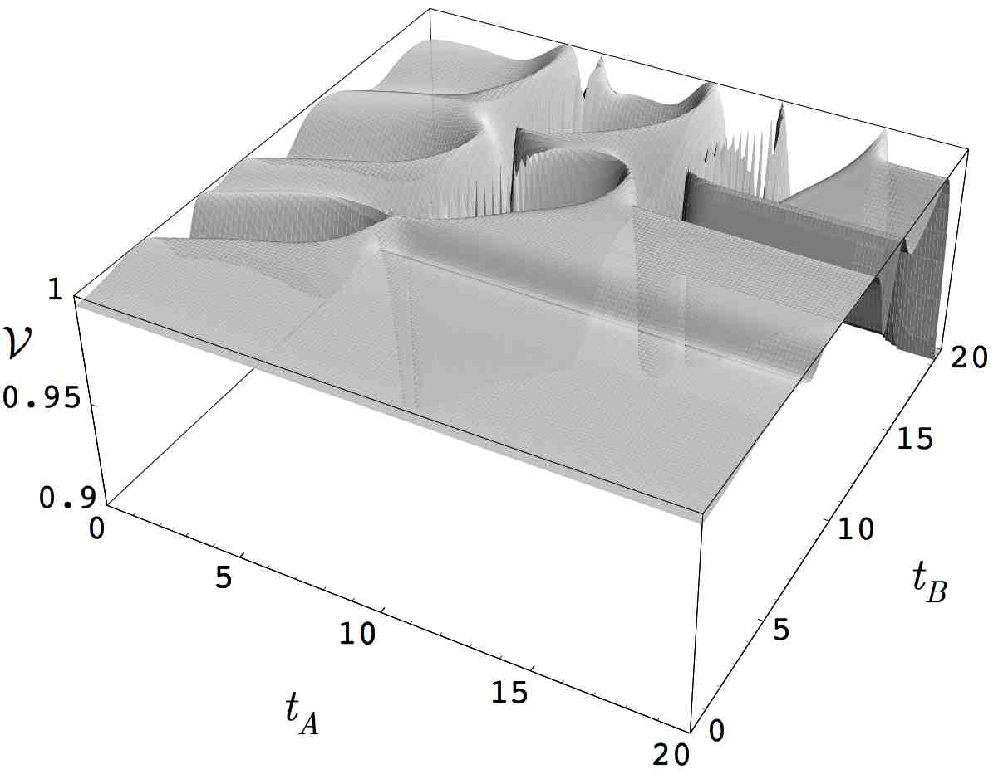}\includegraphics[width=4cm]{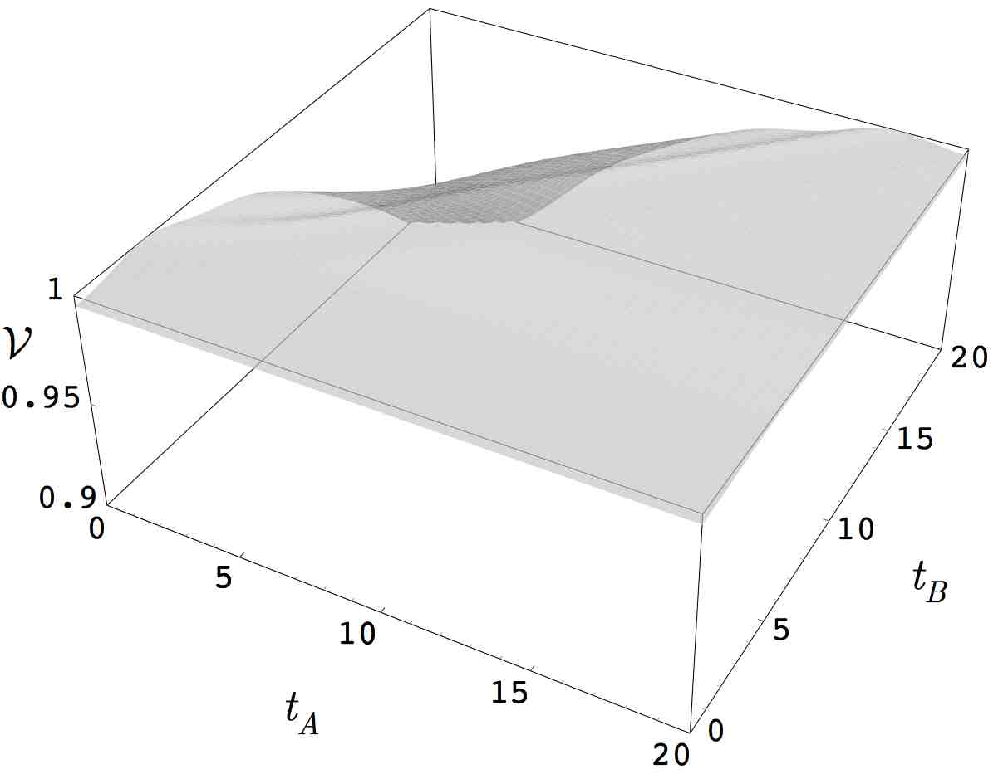}\includegraphics[width=4cm]{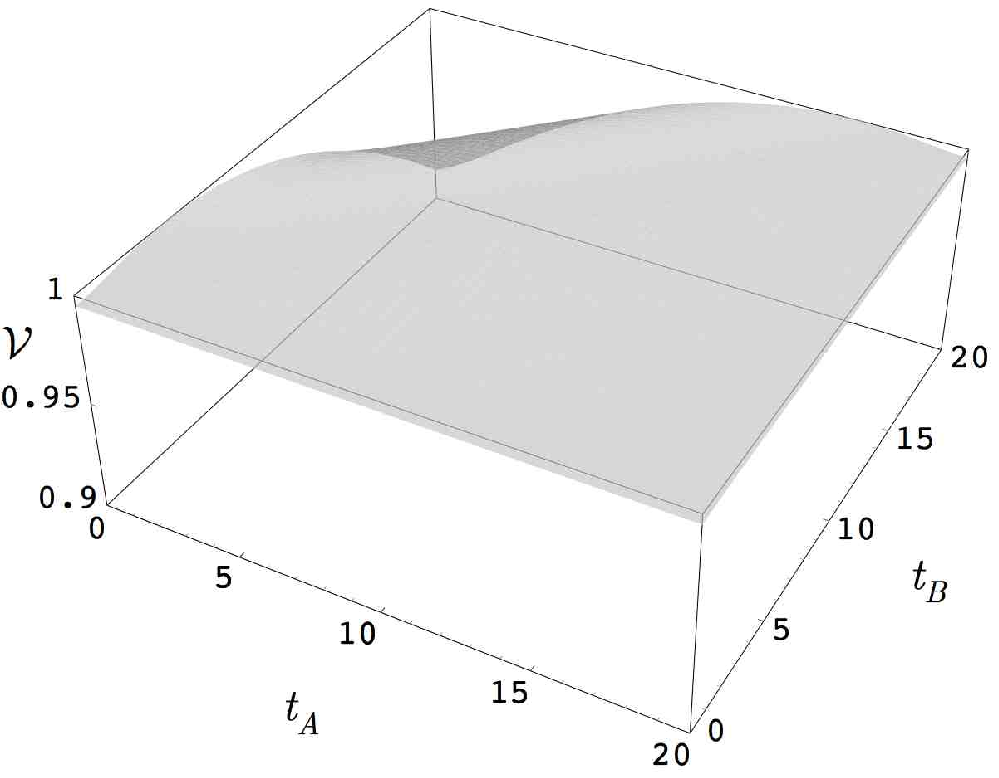}}
\subfigure[]{\includegraphics[width=16cm]{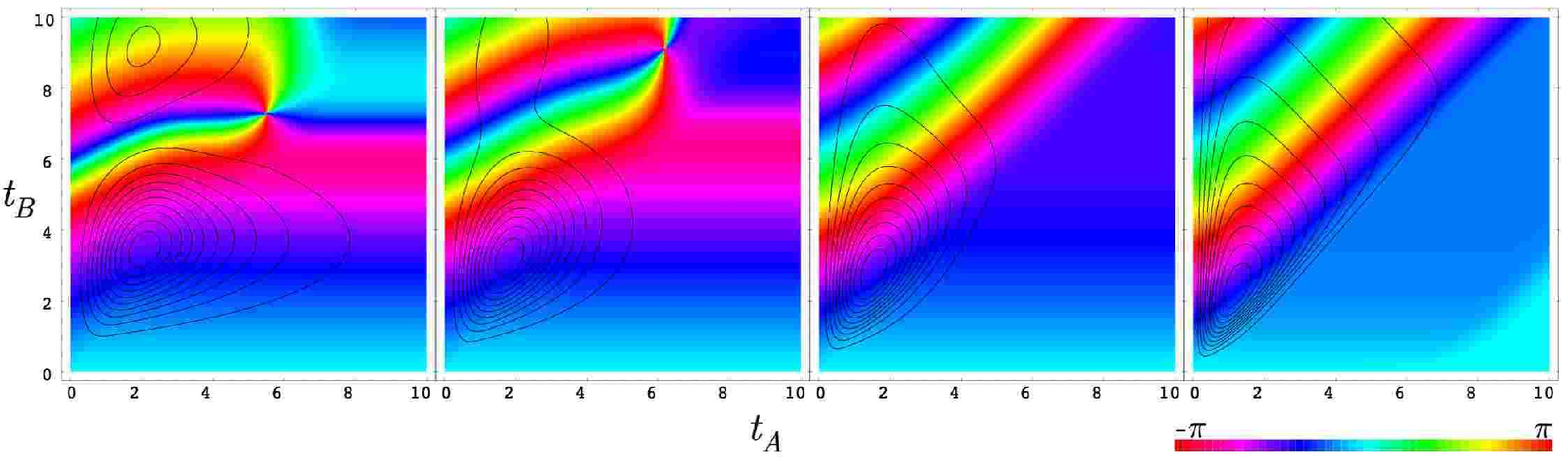}\label{fig:PhaseVariation}}
\caption{
In each row, from left to right, $\kappa=0.5, 1, 2, 4$ and $\Gamma_{d}=\Gamma_{s}=0$ in all figures.  (a) Time dependent probability distribution, $\prob(t_A ,t_B)$, for a balanced system. 
(b) Visibility, $\vis(t_A,t_B)$, for an unbalanced system, where $q_{1,2,3}=1$ and $q_4=1.1$.  (c)  Relative phase,  $\varphi(t_A,t_B)$, for a balanced system.  Superimposed on each panel is a contour plot of $\prob$.}
\end{figure*}

In this section, we present the results of computations for $\meanvis$ for unbalanced systems, and results for $\Q$ which characterizes the extent to which Bell-inequality violations may be observed.  For the problem parameters, we take  experimentally relevant values typical for GaAs self-assembled dots  $\Delta=50$  $\mu$eV \cite{fli01}, $\xi=0.5$ meV, $q=50$ $\mu$eV \cite{ima99, ray02}, $\kappa>100 \mu$eV, though values of $\kappa$ much lower than this may be possible with novel hemsipherical cavities \cite{ray02}.  Throughout  this section we rescale all energies so that $q_i=1$, $\Delta=1$, $\xi=10$.  Time is also rescaled accordingly, so one time unit corresponds to 83 ps.


Figure \ref{fig:BalancedProb} shows plots of the probability distribution of emission times for a balanced cavity with no leakage channels.  Numerically computed visibility is unity to within numerical accuracy and $\meanvis=1$ to within $10^{-4}$, when integrating out to $t_A=t_B=200$.  Notice Rabi oscillations in emission time for strong coupling ($\kappa<q_i=1$) and exponential decay for weak coupling ($\kappa>q$).   
For strong coupling there is a significant probability of emitting photons in either order, but in weak coupling the order $t_B>t_A$ is strongly favoured, indicated by the sharp edge along $t_A=t_B$.

We also note that in the weak coupling regime, $\prob$ has a tendency to broaden with increasing $\kappa$, which is somewhat counter-intuitive, since larger $\kappa$ corresponds to a more leaky cavity, and one would expect the photon-component of the internal state to leak away more rapidly.  However, this phenomenon may also be seen in the much simpler case of a single two-level atom interacting with coupling rate $q$ with a single optical mode of a leaky cavity.  In that case it is straightforward to show that there is in eigenvalue of the effective Hamiltonian for the open system given by $q^2 \kappa^{-1}/2+O(\kappa^{-2})$, which corresponds to a long time constant for large $\kappa$.  When $\kappa\approx q$ there is a kind of impedance matching, and the temporal extent of $\prob$ is smallest.

\begin{figure*}
\subfigure[]{\includegraphics[height=6cm]{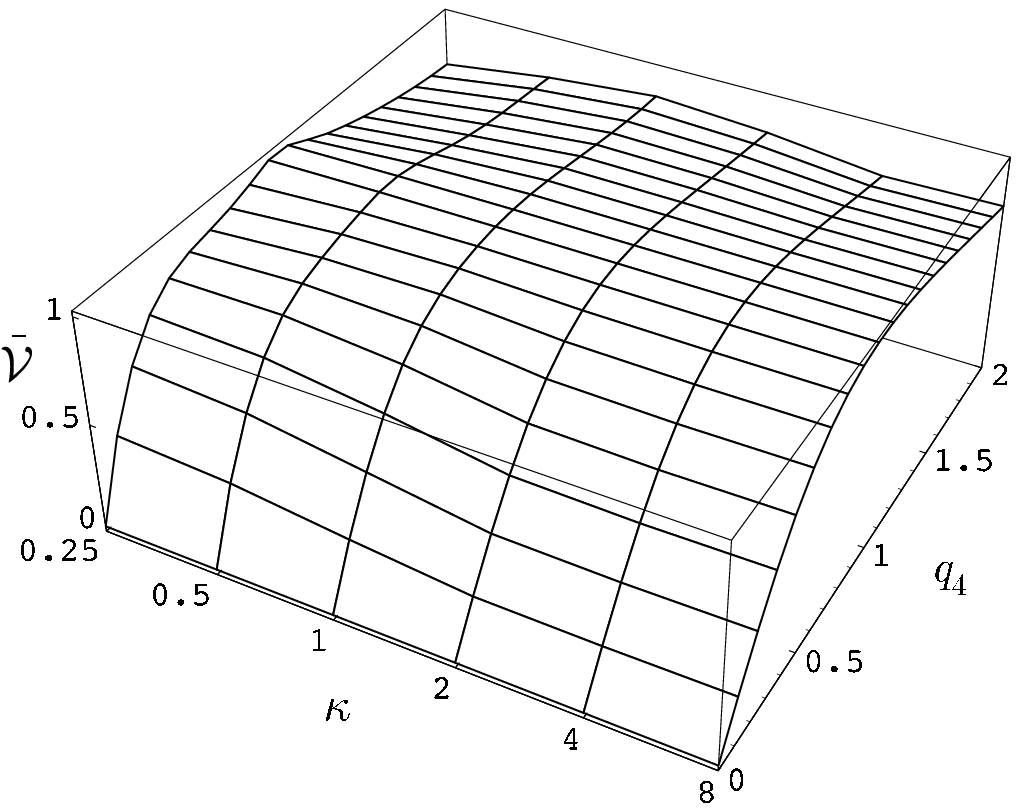}\label{fig:MeanVis_q4}}\hfill
\subfigure[]{\includegraphics[height=6cm]{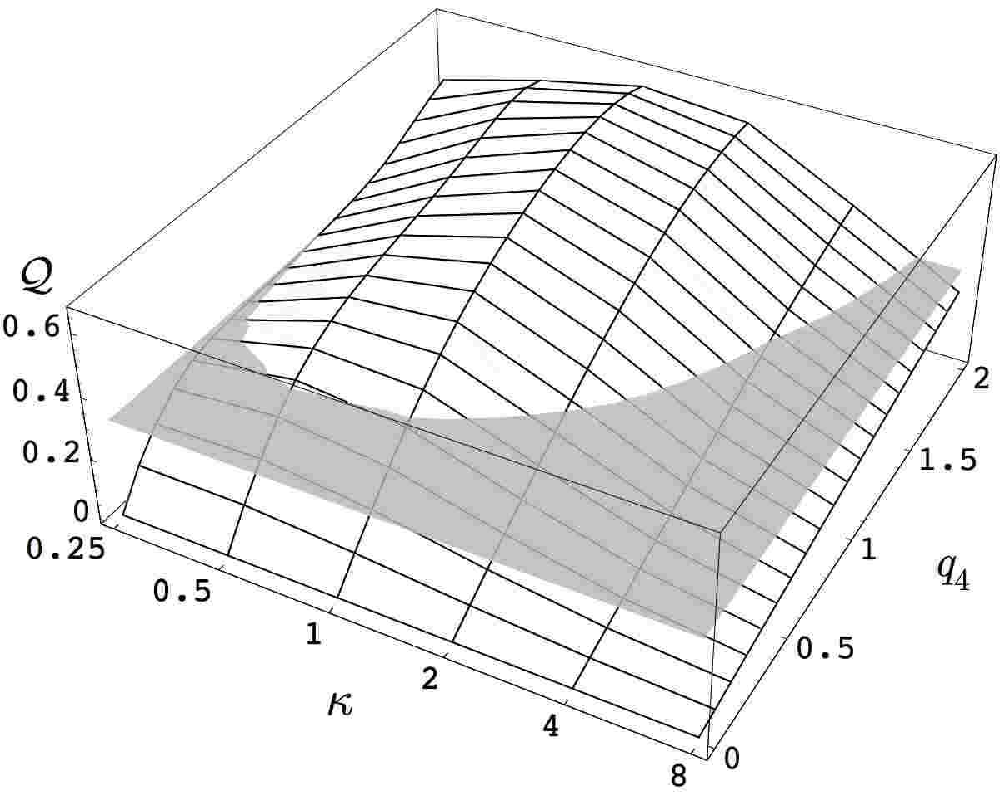}\label{fig:Q_q4}}\\
\subfigure[]{\includegraphics[height=6cm]{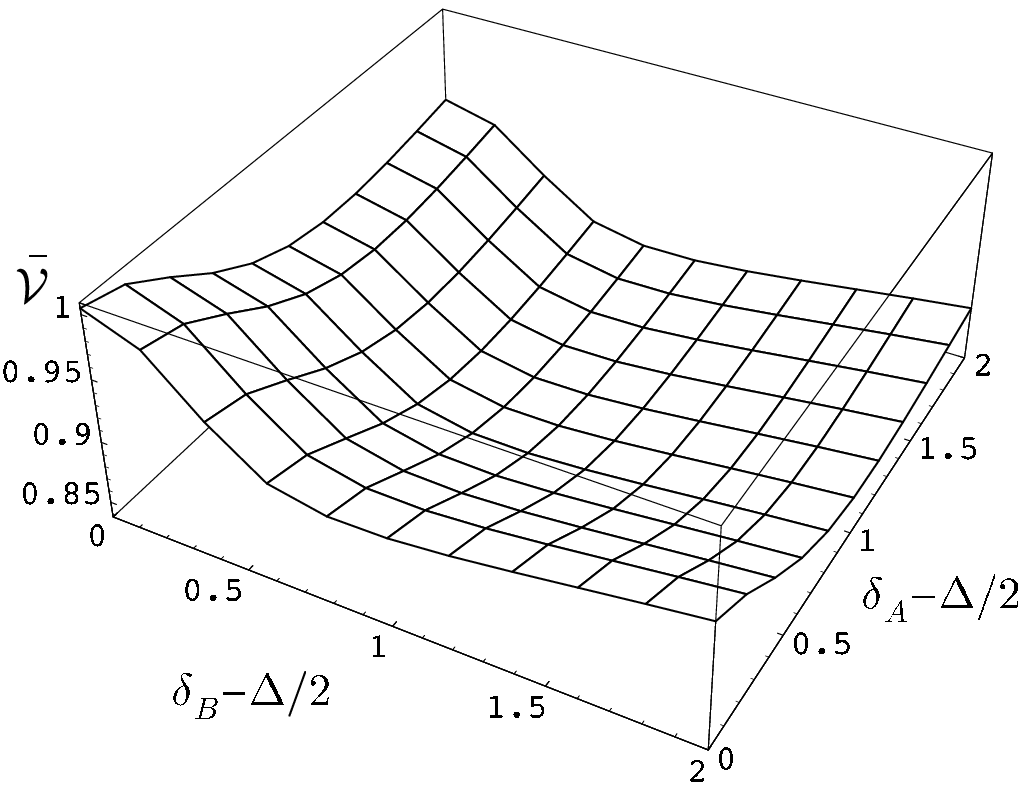}\label{fig:MeanVis_dAdB}}\hfill
\subfigure[]{\includegraphics[height=6cm]{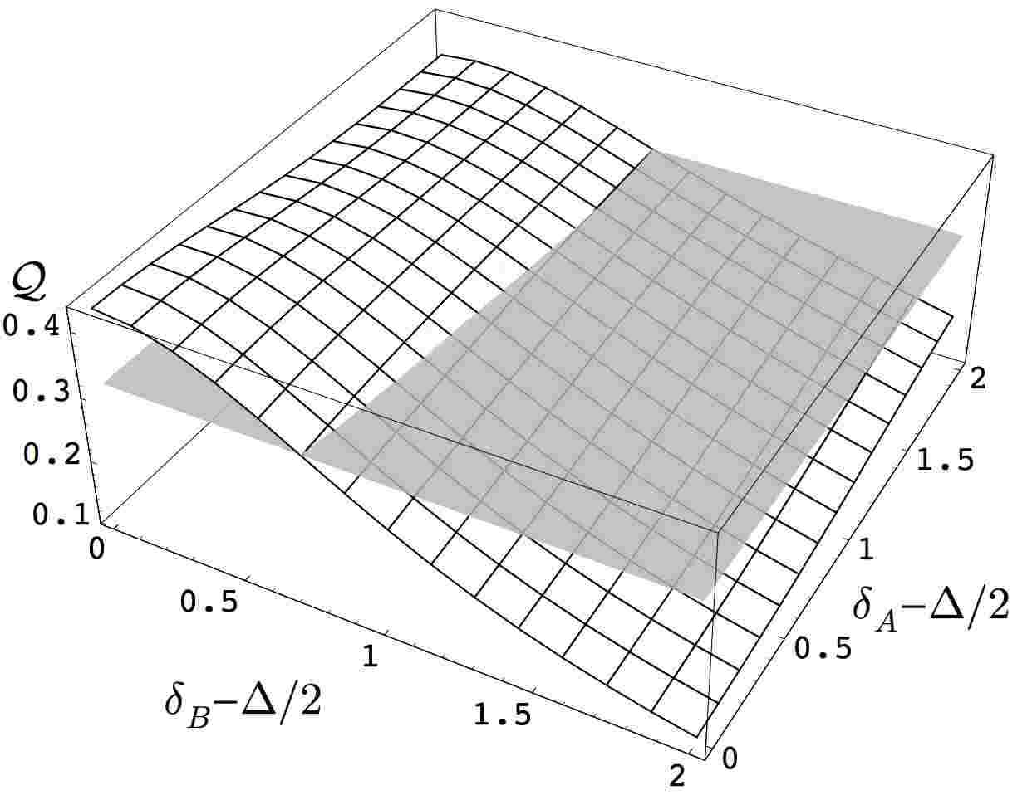}\label{fig:Q_dAdB}}
\caption{\label{fig:MeanVisQ}
Dependence of (a) $\meanvis$ and (b) $\Q$  on $q_4$ for various $\kappa$. Dependence of (c) $\meanvis$ and (d) $\Q$ on $\delta_A$ and $\delta_B$.  $\meanvis$ and $\Q$ are not sensitive to the sign of $\delta_A$ or $\delta_B$ so other quadrants look similar and are not displayed.  The grey plane at $\Q=0.316$ demarks the threshold above which Bell-inequality violations may be observed.}
\end{figure*}


As established previously, the visibility is unity for a balanced system.  For an unbalanced system the visibility drops below unity, as shown in  \fig{fig:UnbalancedVis} where $q_4=1.1$  (with $q_{1,2,3}=1$), for different values of $\kappa$.   The probability density $\prob$ for this case looks very similar to \ref{fig:BalancedProb} so is not shown here.  We note that the visibility depends only on $t_{B}$ when $t_{A}>t_{B}$, i.e. it is frozen at the value it reaches at $t_{B}$.  Notice that the probability, $\prob$, of emitting a photon pair  is small at the the same times that $\vis$ has large excursions from unity, which means $\meanvis$ is not affected as much as one might expect given the large fluctuations in $\vis$.  The difference between strong and weak coupling is striking, again with oscillations being replaced by decay.


The relative phase, $\varphi$, is shown in \fig{fig:PhaseVariation} for a balanced system.  Also superimposed on each panel is a contour plot of the emission probability density, $\prob$.  In the strong coupling regime, during Rabi oscillation peaks, the phase accumulates relatively slowly, with rapid phase rotations in between.  
In the weak coupling regime, the phase accumulates at a fairly constant rate, which is roughly proportional to $\Delta$.  The diagonal stripes indicates that in weakly coupled cavities the phase accumulation depends  only on the time interval between photon emission, $t_B-t_A$, in contrast to the much more complicated dependence of the phase in the strong coupling regime, which shows phase singularities.


Figure \ref{fig:MeanVis_q4} shows the variation of $\meanvis$ versus $q_4$ for various $\kappa$.  For $q_4=0$ one decay path is turned off so we expect completely non-entangled photon pairs, and this is evident in \fig{fig:MeanVis_q4} as  $\meanvis=0$ when $q_4=0$.  We also expect that $\meanvis=1$ when $q_4=1$, since then the couplings are again balanced, and this also is evident in \fig{fig:MeanVis_q4}.  The variation with $q_2$ is identical to that displayed here, whilst the variation of $\meanvis$ with $q_1$ and $q_3$ is qualitatively very similar, so is not shown here.

As discussed earlier, $\Q$ is a significant quantity which determines whether the photon pair can produce Bell-inequality violations in the absence of time-resolved detection, so that phase is ignored.  In particular, as shown in \fig{fig:BellViolation} if $\Q>0.316$ then the photon pair can produce Bell-inequality violations, even in the case that $\varphi$ is ignored.  Figure \ref{fig:Q_q4} shows $\Q$ for the same values of $q_4$ and $\kappa$ as in \fig{fig:MeanVis_q4}, where the grey plane demarks the threshold, $\Q=0.316$, to see Bell-inequality violation, as it will in all following plots of $\Q$.  Whilst $\Q$ is everywhere less than unity, there are parameter values where Bell-inequality violation may still be observed without using time-resolved detection.

If the cavity geometry is  such that $\delta_A\neq\Delta/2$ or $\delta_B\neq\Delta/2$, then the cavity is unbalanced.  This has the effect of reducing $\meanvis$ below unity, as shown in \fig{fig:MeanVis_dAdB}.  Results are only displayed for $\delta_{A,B}>\Delta/2$, but the other quadrants are similar.  Figure \ref{fig:Q_dAdB} shows $\Q$ for the same parameter values, again with the plane denoting the threshold for Bell-inequality violation.

\begin{figure}
\includegraphics[height=6cm]{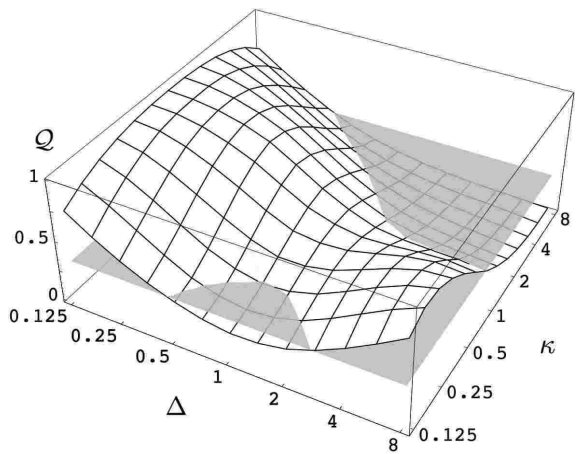}
\caption{\label{fig:Q_kappa_delta}
$\Q$ versus $\Delta$ for $\kappa$.  Varying $\Delta$ changes the rate of phase accumulation between the photon detection events.  Grey plane as in \fig{fig:MeanVisQ}.}
\end{figure}

The phase accumulates in between the photon detection events roughly at a rate proportional to $\Delta$, the splitting due to dot asymmetry, and this directly affects $\Q$, since for smaller $\Delta$ we expect the phase to be more nearly constant over the photon emission lifetime.  This may be seen in \fig{fig:Q_kappa_delta}, where for small $\Delta$, $\Q$ approaches unity, although Bell-inequality violation may still be seen for a wide range of $\kappa$ and $\Delta$.


\begin{figure*}
\subfigure[]{\includegraphics[height=6cm]{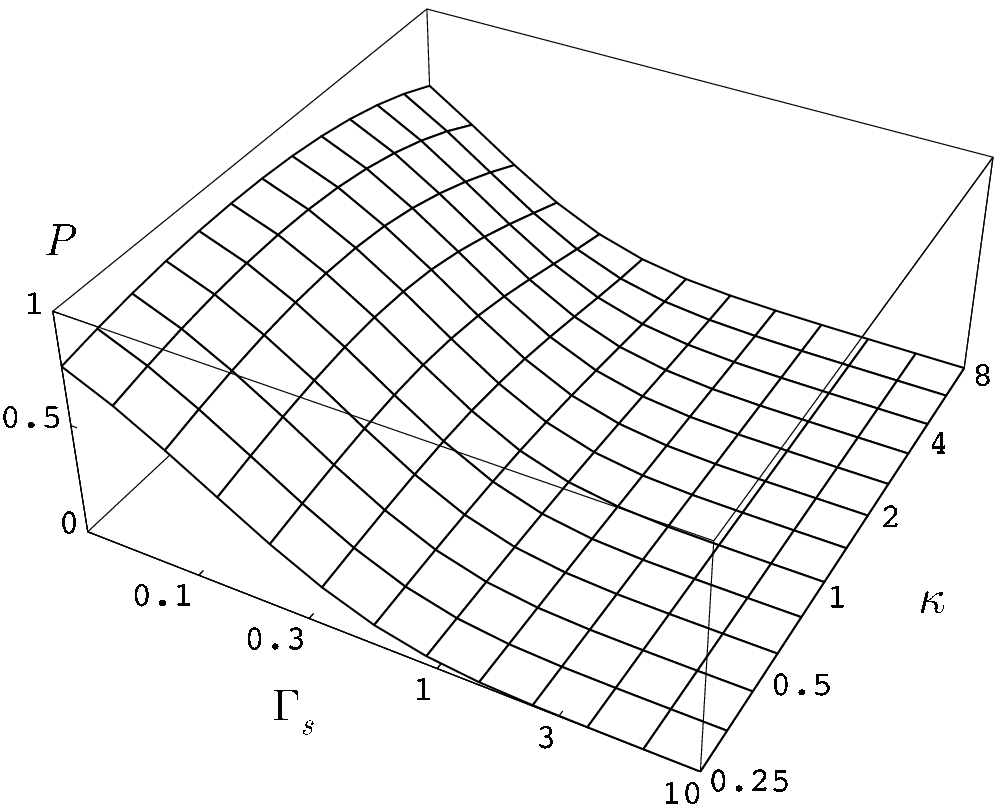}\label{fig:P_spont}}\hfill
\subfigure[]{\includegraphics[height=6cm]{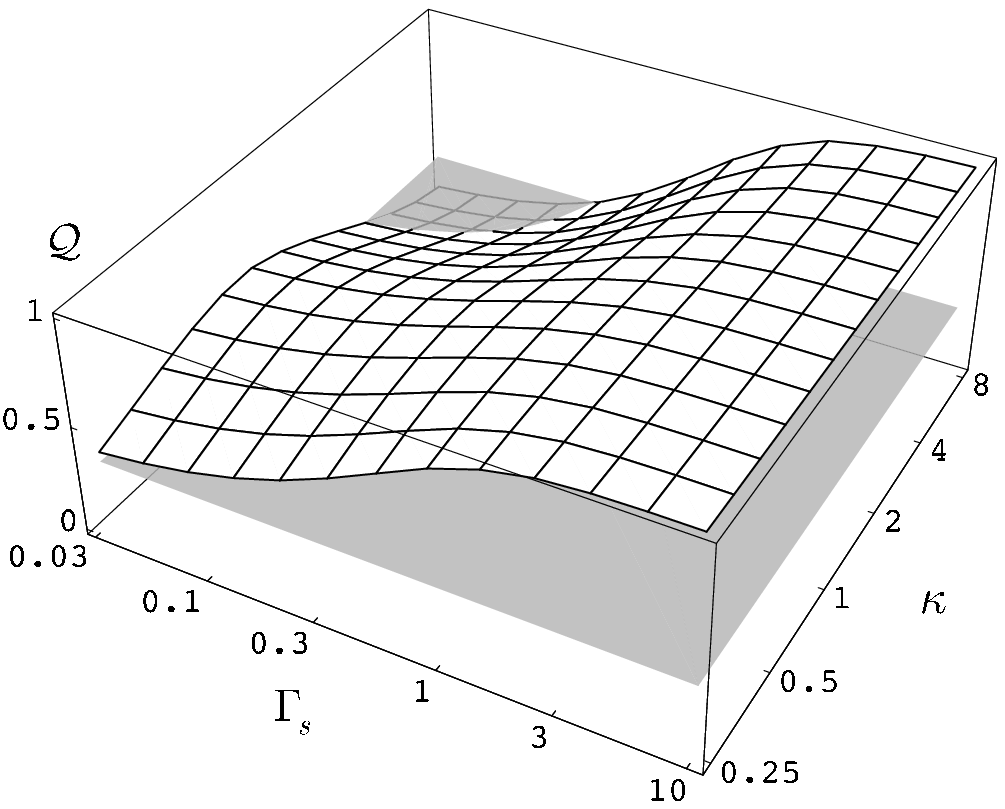}\label{fig:Q_spont}}
\subfigure[]{\includegraphics[height=6cm]{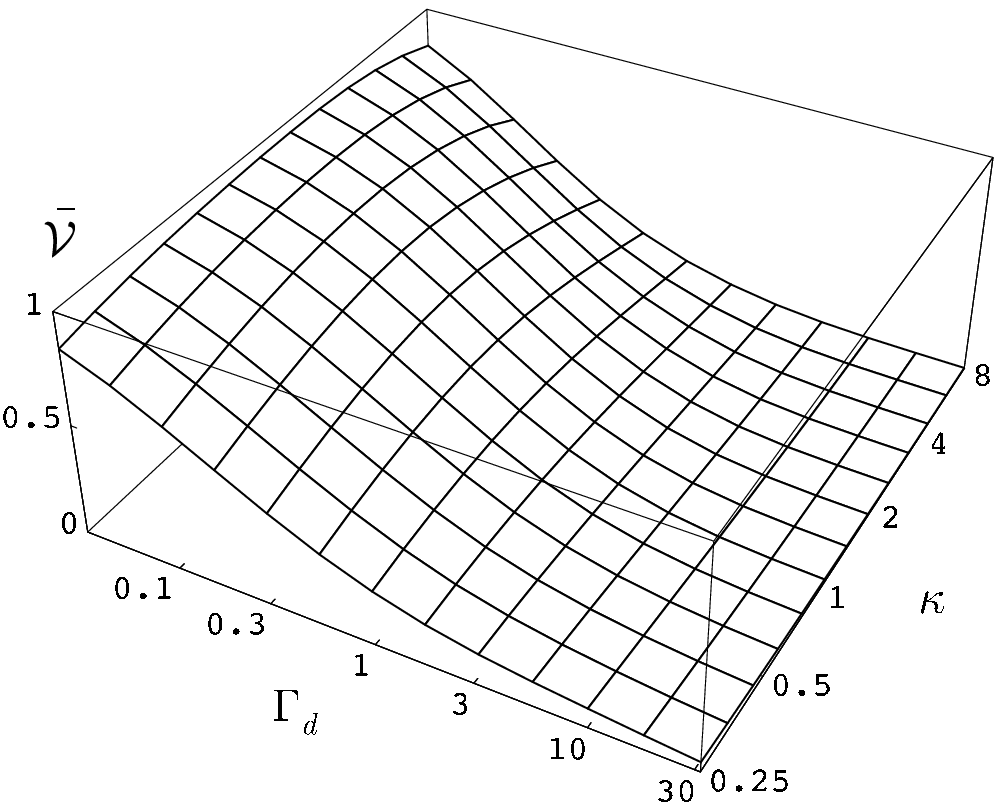}\label{fig:MeanVis_deph}}\hfill
\subfigure[]{\includegraphics[height=6cm]{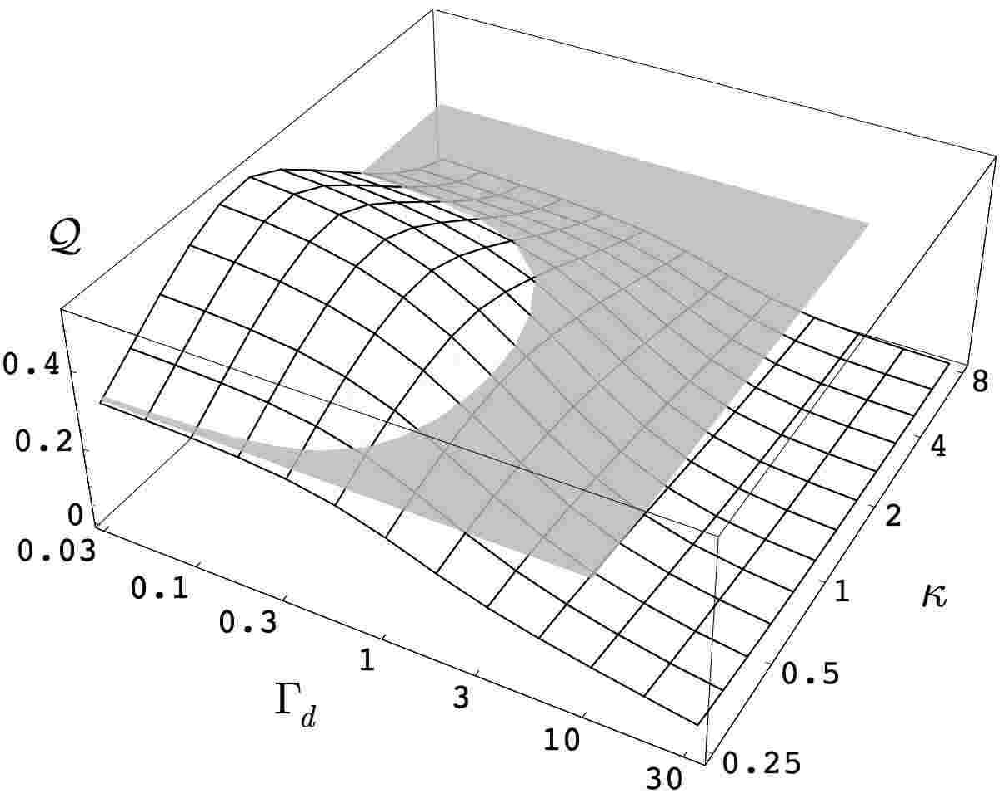}\label{fig:Q_deph}}
\caption{\label{fig:leakage}
(a) $P$ and (b) $\Q$ versus spontaneous emission $\Gamma_s$, for different values of $\kappa$.  (c) $\meanvis$ and (d) $\Q$ versus dephasing rate $\Gamma_d$ for various $\kappa$.  Grey plane as in \fig{fig:MeanVisQ}.}
\end{figure*}

The two different leakage channels that we consider in this paper are spontaneous emission into non-cavity modes, which occurs at a rate $\Gamma_s$, and dephasing which happens at a rate $\Gamma_d$. 

 The spontaneous emission does not affect the visibility of those photon pairs that arrive at the detector, but it does change the rate of detection, since some photons are lost.  The reduction in photon detection rate, given by $P$, is shown in \fig{fig:P_spont}.  The roll-off in $P$ is roughly  proportional to $\Gamma_s^{-1}$.  Surprisingly, the spontaneous emission enhances $\Q$, which may be seen in  \fig{fig:Q_spont}, although we note that the source is then no longer deterministic.  Experimental work indicates that high $\beta$-factors, up to $\beta=0.9$, are possible \cite{ger98} and recently $\beta=0.83$ has been observed \cite{pel02}, which is the fraction of photons emitted into the cavity mode, corresponding to the fraction of photons emitted into the desired cavity modes.  Interpreting $P$ as the $\beta$-factor, from \fig{fig:P_spont} we surmise that for $P\approx0.9$, the experimentally relevant range of the spontaneous emission rate is $\Gamma_s\ll0.1=5$ $\mu$eV, which is a regime in which spontaneous emission is negligible.

Figures \ref{fig:MeanVis_deph} and \ref{fig:Q_deph} shows the effect of the phenomenological dephasing term, $\Gamma_d$, for different values of $\kappa$.  $\meanvis$  and $\Q$ decay roughly as $\Gamma_d^{-1}$.  In all panels of \fig{fig:leakage}, there is a peak along $\kappa\approx q$ which is due to the fact that $\prob$ is temporally narrowest when this condition is met, and hence there is less time for leakage to take place.  For very cold temperatures, around 1 K or lower, pure dephasing rates have been observed to be around 1$\mu$eV \cite{bir01}, corresponding to $\Gamma_d=0.02$, which is negligible.  For higher temperatures, the pure dephasing has been observed to increase at roughly 0.5 -- 1.6 $\mu$eV/K \cite{bir01,bor99}.  From    \fig{fig:Q_deph}, the pure dephasing becomes important near $\Gamma_d\lesssim1$ corresponding to a temperature between 30 and 100 K for the experimentally relevant range given above.

\section{Discussion}

In the previous section we found that the numerical results for a balanced system concur with the analytic result derived in Section \ref{sec:analytic_result} where we established that the visibility is unity in this case.  We also noted that $\meanvis$ is degraded by any effect which may cause the cavity or couplings to be unbalanced.  Imperfections in the cavity geometry will result in an unbalanced cavity so $\delta_{A,B}\neq\Delta/2$, and it was shown above that this reduces $\meanvis$.  Similarly unbalanced coupling constants also  results in decreased visibility.  

\begin{figure*}
\subfigure[]{\includegraphics{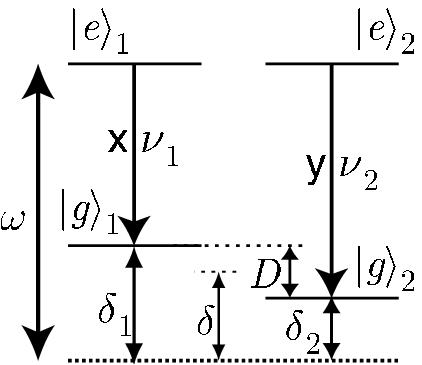}\label{fig:ITLS}}\hfill
\subfigure[]{\includegraphics[width=5.5cm]{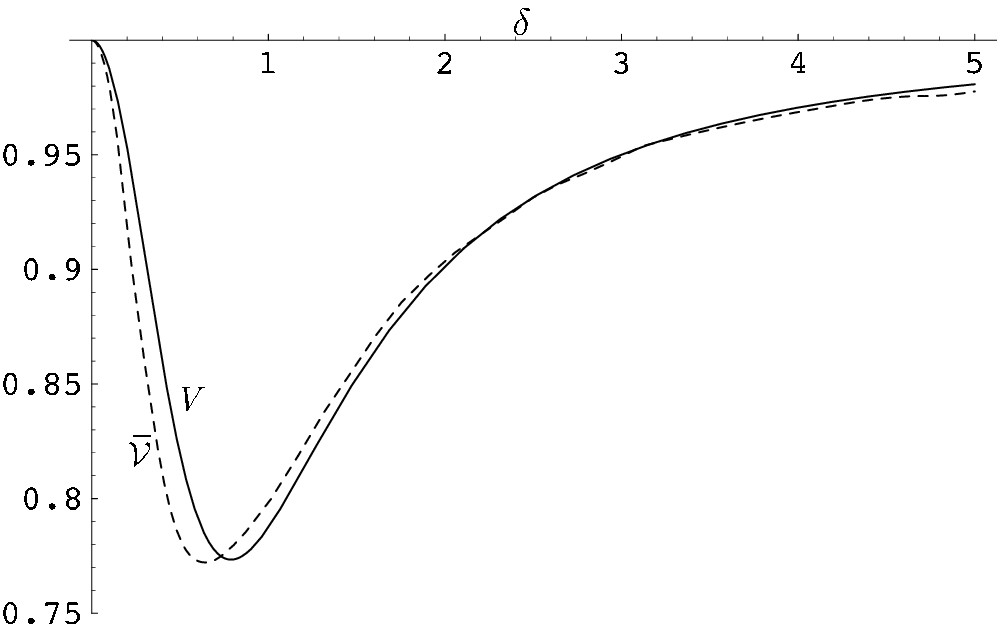}\label{fig:PopulationRatioDetuning}}\hfill
\subfigure[]{\includegraphics[width=5.5cm]{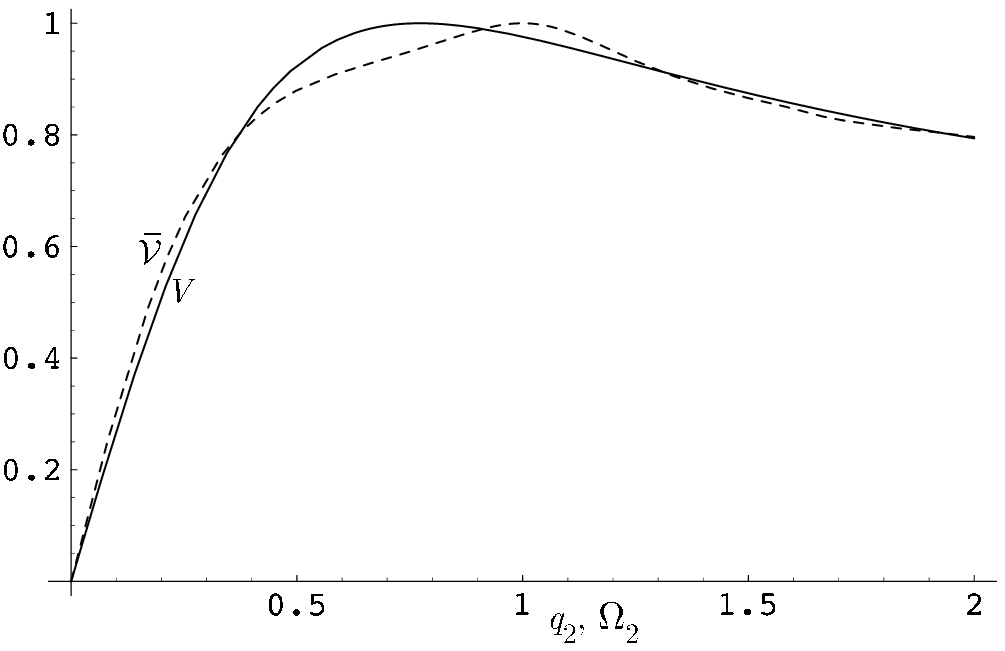}\label{fig:PopulationCouping}}
\caption{\label{fig:PopulationRatio} 
(a) Energy levels for independent two-level system. (b) Equation (\ref{eqn:PopulationRatioDetuning}) (solid) versus $\delta$ and also $\meanvis$ (dotted) versus $\delta_B$.  (c) Equation (\ref{eqn:PopulationRatioCoupling}) (solid) versus $\delta$ and also $\meanvis$ (dotted) versus $\delta_B$.  Parameters as usual except $\kappa=0.4$.}
\end{figure*}

Both of these effects may be understood heuristically using a much simpler model which captures the gross features seen in \figs{fig:UnbalancedVis} and \ref{fig:MeanVis_dAdB}.  Firstly, we note that a two-photon state given by $\alpha_\x\ket{\x\x}+\alpha_\y\ket{\y\y}$ will produce two-photon interference fringes with visibility 
\begin{equation}
V=\frac{2|\alpha_x \alpha_\y|}{|\alpha_\x|^2+ |\alpha_\y|^2}.
\end{equation}
  Secondly, we make two \emph{ad hoc} simplifications of the level structure of the quantum-dot shown in  \fig{fig:EnergyLevels}.  These simplifications are (i) to ignore the crystal ground state, $\ket{\g}$, and the corresponding transitions thereto, and (ii) to treat the remaining three level system, composed of $\ket{\bx}$, $\ket{\xx}$ and $\ket{\xy}$ as a pair of independent two-level systems (TLS), $\{\ket{g}_1,\ket{e}_1\}$ and $\{\ket{g}_2,\ket{e}_2\}$, each of which interact with one of a pair of degenerate cavity modes distinguished by polarisation.  With these two assumptions, the energy level structure becomes that pictured in \fig{fig:ITLS}.

The physical motivation for these seemingly arbitrary assumptions is firstly that once the biexciton decay proceeds along the $\x$- or $\y$-polarised paths of  \fig{fig:EnergyLevels}, the resulting two-photon amplitudes, $\alpha_{\x,\y}$ are determined, even though the dynamics of the emission are not complete.  Thus the two-photon amplitudes are largely determined by the initial single-photon decay process, justifying (i).  Secondly, whilst the sum of probabilities to take the $\x$- or $\y$-polarisation decay paths is unity, apart from this constraint the rate equations for the two decay processes are otherwise uncoupled, so the system similar to a pair of uncoupled TLS', one for each decay path, justifying (ii).  Ultimately, this highly simplified model will be verified by its qualitative agreement with the more realistic model discussed throughout this paper, and its value is in the intuition it lends about the origin of the effects seen in the numerical calculations.

A TLS interacting with a cavity mode is well understood in terms of the Jaynes-Cummings (JC) model \cite{yamamoto}.  For TLS' initially in the state $\ket{e}_i$ ($i=1,2$ is the TLS label) with energy spacings $\nu_i$, oscillator frequency $\omega_i$ and detuning $\delta_i=\nu_i-\omega_i$, with TLS-cavity coupling rate $\Omega_i$, (see \fig{fig:ITLS}), the time-averaged photon population is given by $p_i=\Omega_i^2/(2R_i^2)$, where $R_i=\sqrt{\delta_i^2+\Omega_i^2}$, and we conclude that the average amplitude of photon occupation satisfies  
\begin{equation}
|\alpha_{\x,\y}|=\frac{\Omega_{1,2}}{\sqrt{2}R_{1,2}}.\label{eq:alpha}
\end{equation}

 
We now compare the predictions of this simple model with the more complete one for  an unbalanced cavity, wherein the cavity mode is not tuned to the mean of the transition frequencies, $\omega\neq (\nu_1+\nu_2)/2$.  If the TLS'  are detuned by an amount $\delta_{1,2}=\delta\mp D/2$ respectively from the degenerate cavity modes, (see \fig{fig:ITLS}), each with the same cavity-coupling strength $\Omega_{1,2}=\Omega$, the visibility is then given by
\begin{equation}\label{eqn:PopulationRatioDetuning}
V=\frac{2 R_1 R_2}{R_1^2+R_2^2},
\end{equation}
where we have taken $|\alpha_{\x,\y}|$ from \eqn{eq:alpha}.  This expression is plotted in \fig{fig:PopulationRatioDetuning} as a function of $\delta$ (using $\Omega=0.61, D=1$) along with $\meanvis$ (using $q_{1,2,3,4}=1, \kappa=0.4, \Delta=1, \delta_A=\Delta/2, \xi=10, \Gamma_{d,s}=0$).  Clearly the forms of the two traces are in qualitative agreement demonstrating the heuristic validity of the simple model.  The value $\Omega=0.61$ is selected  to fit \eqn{eqn:PopulationRatioDetuning} to the numerically computed $\meanvis$, but it is of the same order as $q_i=1$.

We also compare the predictions of the simple model for unbalanced coupling to the realistic model, and so we take  $\Omega_1\neq\Omega_2$, but suppose the detunings between  the two-level systems and their respective harmonic oscillators are equal, $\delta_1=-\delta_2=\delta$.  It is straightforward to show that
\begin{equation}\label{eqn:PopulationRatioCoupling}
V=\frac{2 R_1\Omega_1 R_2\Omega_2}{(R_1\Omega_1)^2+(R_2\Omega_2)^2},
\end{equation}
which is plotted in \fig{fig:PopulationCouping} (using $\Omega_1=0.42, \delta=1.69$- both fitted parameters), along with $\meanvis$ (other parameter values as in \fig{fig:PopulationRatioDetuning}).

The simple, heuristic model of two uncoupled two-level systems predicts a visibility, $V$, that is qualitatively in agreement with $\meanvis$ calculated using the complete model discussed in earlier sections.  Thus we can understand the most significant effect of variation of  $q_i$ and $\delta_i$ on $\meanvis$ is to change the relative amplitudes to take each of the two decay paths illustrated in \fig{fig:EnergyLevels}.  Since the photon pair is only maximally entangled when the amplitudes of the $\ket{\x\x}$ and $\ket{\y\y}$ components are equal in magnitude (i.e. for the state $(\ket{\x\x}+e^{-i \phi}\ket{\y\y})/\sqrt{2}$),  parameter variations that result in unequal decay path amplitudes result in sub-maximally entangled photon pairs.  Such parameter variations correspond directly to the situation of an unbalanced system.

The analysis above gives us some further insight into the decay process.  The maximum amplitude of the  photon excitation is $\Omega^2/R^2$ and so the leakage rate of photons from the cavity will be suppressed by this factor.  That is, we expect that the rate of decay of excitation from the cavity will be roughly $\kappa \Omega^2/(\Omega^2+\delta^2)$.  Therefore, as the detuning $\delta$ increases, the photon emission rate slows roughly as $\sim1/\delta^2$ for $\delta>\Omega$.  This will mean that for detunings significantly larger than the coupling strength, leakage effects will become significant -- the lifetime of the excitation in the cavity will become comparable to the decay rate for dephasing or spontaneous emission.

It is also worth noting that when $\delta\sim2\xi$ (i.e. the exciton-cavity detuning is near the biexciton shift) the model developed in section \ref{sec:Hamiltonian} breaks down, since significant cross coupling between exciton states and cavity modes will set in.

Spontaneous emission decreases the detection probability, which could be corrected with post-selection, since only events in which two-photons are registered count towards the measurement, and as mentioned previously, experimental work has shown that this is negligible for experimentally relevant  systems \cite{ger98}.  Dephasing of intermediate states decreases the visibility exponentially in time.  Temperatures of a few Kelvin provide sufficiently low dephasing rates that it is negligible, but the excitonic dephasing becomes important at temperatures of several tens of Kelvin, \cite{bir01,bor99}.  In principle, these effects may be distinguished using sufficiently fast time-resolved spectroscopy, since spontaneous emission will result in fewer photons reaching the detector, whereas dephasing would result in a time dependent visibility that degrades exponentially with time.

So far we have not addressed the issue of how to experimentally construct a cavity with the required spectrum, shown in \fig{fig:EnergyLevels}, and a detailed proposal for its implementation is beyond the scope of this paper.  The enhanced exciton emission into the cavity mode is known as the Purcell effect and requires small cavity volumes, so that the exciton-cavity mode coupling strength is large and the density of available photon modes is small \cite{kim94,ger98}.  Thus small cavities are necessary, and high Purcell factors have been demonstrated experimentally in single-wavelength sized cavities \cite{ger98,rob01}.

In contrast to the need for small cavities is the relatively small biexciton shift, $2\xi$, which is around 1 meV.  In order for a single Fabry-Per\'ot  resonator to accommodate modes spaced by 1 meV (i.e. the free spectral range, FSR), the cavity length would need to be of the order of $100 \mu$m or more.   This is too long for several reasons, because the Purcell factor decreases with cavity length, and also because growth of such a large heterostructure would be prohibitively difficult.  As a result the cavity to which we have been referring throughout this paper would need be based on a more complicated geometry than merely a pair of distributed Bragg reflectors (DBR) forming a linear resonator.  

We stress that a more complex geometry is not just a requirement of this proposal, but that it would be necessary for a system even with symmetric quantum dots.  If the cavity did not have separate modes near the exciton and biexciton doublet frequencies then only one transition could couple strongly to the cavity, and the other transition would be sufficiently off resonance ($\delta>q$) that the Purcell effect for that frequency would be suppressed, i.e. either the biexciton-exciton or exciton-ground transitions may be well coupled to the cavity, but not both.

It may be possible to engineer a small cavity with a pair of closely spaced modes using photonic crystals.  If, during the growth of each DBR stack one layer was permitted to grow to be larger than $\lambda/4$, then the cavity would look more like two coupled cavities, which may have the desired split modes.  Certainly, geometric effects in micropillars have been shown to produce a pair of modes spaced by $\sim5$ meV \cite{rob01}, though this was due to lifting polarisation degeneracy with elliptical cross-section cavities, which is undesirable for our scheme.

Experiments using confocal hemispherical cavities, consisting of a planar Bragg reflector at the focal plane of a hemispherical reflector, of length 50 to 1000 $\mu$m are currently underway for quantum information processing purposes \cite{ray02}.  In this configuration the cavity mode waist diameter is of comparable size to the optical wavelength and coincident with a quantum dot so that the exciton-cavity mode coupling strength is reasonably large.  This arrangement may provide the two requirements of the present paper: both strong coupling between the dot excitations and the cavity mode (up to several tens of $\mu$eV) and small FSR so that each doublet is on resonance with a nearby mode.  It is quite plausible that by tuning the cavity length to vary the FSR and applying an external DC electric field to induce a Stark shift in the doublet frequencies, one may bring both doublets close to cavity modes simultaneously, as depicted in \fig{fig:Spectrum}, thereby realising the requirements of this proposal.

\section{Summary}

We have shown analytically that using a cavity with a particular mode structure facilitates the production of polarisation entangled photon pairs from an asymmetric  quantum dot, which otherwise produces photon pairs entangled in both polarisation and frequency.  We demonstrated this by computing the visibility of two-photon interference fringes produced using photons generated from such a cavity-quantum dot structure, and related this to their potential to demonstrate Bell-inequality violations.

We have quantified the effect of various errors in the cavity mode structure, showing that the visibility is not degraded badly by mistuned cavities or unbalanced dipole coupling strengths, and for experimentally accessible regimes is above the threshold at which Bell-inequality violations may be detected.

Of major significance to this scheme is the phase accumulated between single photon emission events.  By defining the phase-averaged visibility we were able to compute the effect of ignoring this phase on Bell-inequality and two-photon visibility measurements.  Such phase-ignorance arises when the available time-resolution  of the photon detection apparatus is longer than the asymmetry splitting, $\Delta$.  

We showed that ignoring the phase reduces the effective entanglement, but there are still experimentally accessible regions of parameter space that exhibit Bell-inequality violation, even when the phase is ignored, and the two-photon states are thus potentially useful sources of entanglement.

\begin{acknowledgments}
We would like to thank Andrew Shields and Mark Stevenson who prompted this work, and for their helpful discussions and comments.  We would also like to thank Sean Barrett and Michael Raymer for useful conversations and suggestions.  TMS thanks the Hackett Scholarships committee and the CVCP for financial support.  GJM acknowledges financial support from the CMI.  CHWB thanks the EPSRC for financial support.
\end{acknowledgments}

\end{document}